\documentclass[aps,pra,twocolumn,superscriptaddress]{revtex4-2} 

\usepackage[final,authormarkup=none,defaultcolor=purple]{changes}
\definechangesauthor[name={Benedikt}, color=purple]{BT}
\definechangesauthor[name={Anders}, color=purple]{AS}
\definechangesauthor[name={Emil}, color=purple]{EH}
\definechangesauthor[name={Soubhadra}, color=purple]{SM}

\usepackage[]{graphicx}
\usepackage[dvipsnames]{xcolor}
\usepackage[inkscapelatex=false]{svg}
\usepackage[colorlinks=true,citecolor=NavyBlue,linkcolor=RubineRed,urlcolor=Cerulean,pdfencoding=auto]{hyperref}
\usepackage[]{amsmath}
\usepackage[]{amssymb}
\def\enquote#1{``#1''}
\def\bra#1{\mathinner{\langle{#1}|}}
\def\ket#1{\mathinner{|{#1}\rangle}}
\def\braket#1{\mathinner{\langle{#1}\rangle}}

\def\abs#1{\mathinner{|{#1}|}}
\def\abssq#1{\mathinner{|{#1}|}^2}
\def\proj#1{\ket{#1}\bra{#1}}
\def\op#1#2{\ket{#1}\bra{#2}}
\def\comm#1#2{\mathinner{[{#1},{#2}]}}
\DeclareMathOperator{\I}{\openone}
\DeclareMathOperator{\tr}{tr}

\begin{document}

\title{Single and Double-click High-Rate Entanglement Generation Between Distant Ions Using Multiplexed Atomic Ensembles}

\author{Benedikt Tissot}
\email[]{benedikt.tissot@nbi.ku.dk}
\affiliation{Center for Hybrid Quantum Networks (Hy-Q), Niels Bohr Institute, University of Copenhagen, Jagtvej 155A, DK-2200 Copenhagen, Denmark}

\author{Soubhadra Maiti}
\affiliation{QuTech, Delft University of Technology, Lorentzweg 1, 2628 CJ Delft, The Netherlands}
\affiliation{EEMCS, Quantum Computer Science, Delft University of Technology,
Mekelweg 4, 2628 CD Delft, The Netherlands}

\author{Emil R. Hellebek}
\affiliation{Center for Hybrid Quantum Networks (Hy-Q), Niels Bohr Institute, University of Copenhagen, Jagtvej 155A, DK-2200 Copenhagen, Denmark}

\author{Anders Søndberg Sørensen}
\email[]{anders.sorensen@nbi.ku.dk}
\affiliation{Center for Hybrid Quantum Networks (Hy-Q), Niels Bohr Institute, University of Copenhagen, Jagtvej 155A, DK-2200 Copenhagen, Denmark}

\begin{abstract}
In an accompanying paper 
\nocite{LETT}[arxiv:2511.04488], we introduced an approach to interface trapped-ion quantum processors with ensemble-based quantum memories by matching a spontaneous parametric down conversion source to both the ions and the memories.
This enables rapid entanglement generation between single trapped ions separated by distances of hundreds of kilometers.
In this article, we extend the protocol and provide additional details of the analysis.
Particularly, we compare a double-click and single-click approaches for the ion edge nodes. The double-click approach relaxes the phase stability requirement but is strongly affected by finite efficiencies. 
Choosing the optimal protocol thus depends on the access to the phase stabilization as well as the efficienc\replaced[id=BT]{ies}{y} of \added[id=BT]{the} interface\added[id=BT]{s} of the ions and ensemble-based memories.
\end{abstract}

\maketitle

\section{Introduction}

A central challenge in building a quantum internet is high-rate long-range entanglement generation \cite{kimble-2008-quant_inter,wehner-2018-quant_inter}.
In the quantum internet, \deleted[id=BT]{the} (remote) entanglement serves as a central resource for
communication \cite{ekert-1991-quant_crypt_based_bells_theor,acin-2007-devic_indep_secur_quant_crypt,gisin-2007-quant_commun,pirandola-2020-advan_quant_crypt}, enhanced sensing \cite{wasilewski-2010-quant_noise_limit_entan_assis_magnet,cassens-2025-entan_enhan_atomic_gravim}, and distributed computing \cite{cuomo-2020-towar_distr_quant_comput_ecosy}.
Furthermore, long-range entanglement is a cornerstone for fundamental physics experiments, first and foremost the violation of Bell-inequalities \cite{bell-1964-einst_podol_rosen_parad,clauser-1969-propos_exper_to_test_local,hensen-2015-looph_free_bell_inequal_violat,giustina-2015-signif_looph_free_test_bells,storz-2023-looph_free_bell_inequal_violat}.
Remote entanglement generation has been demonstrated in a variety of different physical systems, e.g.,
between atomic ensembles \cite{chou-2007-funct_quant_nodes_entan_distr,yuan-2008-exper_demon_bdcz_quant_repeat_node,yu-2020-entan_two_quant_memor_via}, trapped ions   \cite{moehring-2007-entan_singl_atom_quant_bits_at_distan,stephenson-2020-high_rate_high_fidel_entan,krutyanskiy-2023-entan_trapp_ion_qubit_separ_by_meter}, color centers in solids \cite{bernien-2013-heral_entan_between_solid_state,sipahigil-2016-integ_diamon_nanop_platf_quant_optic_networ,humphreys-2018-deter_deliv_remot_entan_quant_networ}, and rare earth ions \cite{lago-rivera-2021-telec_heral_entan_between_multim,liu-2021-heral_entan_distr_between_two,ruskuc-2025-multip_entan_multi_emitt_quant_networ_nodes},
where usually, photons are used to herald entanglement between remote matter nodes.
Although photons are the natural carriers for long-distance entanglement distribution, they still require finite time to propagate over large distances, and fiber losses present a significant challenge to achieving high-rate, high-fidelity entanglement.

One potential approach to mitigate communication time and fiber losses is the use of multiplexing,
where multiple entanglement generation attempts are stacked within the communication time.
This can also be viewed as optimizing the resource usage, as it helps saturate the use of fiber modes.
Thus, nodes of a quantum network should preferably combine efficient photon interaction or emission capabilities, ideally naturally supporting multiplexing, with long coherence times and the potential to apply local gate operations.
Identifying nodes that integrate these capabilities while also being compatible with telecom frequencies remains an unsolved challenge.
In Ref. \cite{LETT}, we proposed a protocol for a hybrid system comprising trapped ions and spontaneous parametric down-conversion sources combined with multimode ensemble-based memories (SPDC+M).
The hybrid system promises to unite the high-rate multiplexed entanglement generation over long distances of the SPDC+M nodes with the advanced quantum processing capability of trapped ions \cite{bruzewicz-2019-trapp_ion_quant_comput,moses-2023-race_track_trapp_ion_quant_proces}.

We proposed to generate matter-photon entanglement by matching the envelope of the photon field emitted from the SPDC to the ions, resulting in ion-photon entanglement stored within the quantum memory.
The stored photon is then naturally compatible with a high-speed backbone (BB) built from SPDC+M nodes.
As a result, the protocol allows for the generation of entanglement between trapped ions over long distances at rates that are much higher than what can be achieved by direct ion-ion communication using simpler intermediate systems.
In this article, we delve into the additional details of the ion-ion entanglement generation enabled by the hybrid system.
To this end we consider an optical setup in both the entanglement generation and entanglement swapping steps of the protocols, where photons are (re-)emitted from the nodes,\added[id=AS]{ and} combined via a beam-splitter terminating in two detectors\replaced[id=AS]{. U}{, and u}pon the detection of a single (or two) photons, the successful generation (swapping) of entanglement is heralded, see Fig.~\ref{fig:modules}.
Henceforth, we will refer to the photon detection as a click.
Since, approaches using both a single or two clicks are used to herald entanglement generation and swapping \cite{sangouard-2011-quant_repeat_based_atomic_ensem_linear_optic,beukers-2024-remot_entan_protoc_station_qubit},
we will compare the \added[id=BT]{two-}single-click protocol employed in Ref.~\cite{LETT} \added[id=BT]{(which combines two single-click links in a purification step)} with a double-click protocol.
Both protocols rely on a single-click protocol for the long-distance entanglement generation within the BB due to the beneficial scaling with fiber losses \cite{duan-2001-long_distan_quant_commun_with,sangouard-2011-quant_repeat_based_atomic_ensem_linear_optic,beukers-2024-remot_entan_protoc_station_qubit}.
However, within the edge nodes (EN), a single or double-click  protocol is used to herald the entanglement with a single or with two memories in individual rails.
\deleted[id=BT]{
The main difference is the scaling with the intrinsic efficiencies, e.g., the ensemble memory efficiency, and the phase stability requirements imposed on the ions, as we will discuss in detail below.
}

\added[id=BT]{
  In summary, we will show that both protocols unite the trapped ion toolbox with the multiplexing capacity of SPDC+M nodes, and that this multiplexing leads to a speed-up for long-distance ion-ion links.
  Furthermore, the two-single-click protocol features favorable scaling with the memory efficiency compared to the double-click protocol.
  This is due to the entanglement swaps having to succeed simultaneously (one per rail) in the dual-rail ion-photon entanglement of the double-click protocol.
  This makes the double click protocol more sensitive to photon loss.
  On the other hand, the double click protocol intrinsically has a better fidelity, removing the need for a purification step which is an integral part of the two-single-click scheme.
  Furthermore,   
  the double-click protocol relaxes the phase stabilization requirement compared to the two-single-click protocol, where phase stability is required over multiple entanglement generation attempts.
  As a consequence the double-click protocol is advantageous for efficient memories, while the two-single-click protocol is desirable for inefficient setups, if phase stability can be achieved over many entanglement generation attempts.
}

We organize the remainder of this article as follows.
First, we introduce the SPDC+M nodes and the single-click BB entanglement generation in Sec.~\ref{sec:BB}.
In this section, we include the underlying model for photon detection.
Then in Sec.~\ref{sec:proto}, we provide an overview of the two protocols with the corresponding initial ion-photon states used to interface ions and SPDC+M nodes.
We continue by providing a detailed analysis of the protocols,
the two-single-click protocol in Sec.~\ref{sec:TSC_ana}
and the double-click in Sec.~\ref{sec:DC_ana}.
Based on the detailed analysis, we then compare the protocols with each other, as well as entanglement generation between ions without the SPDC+M BB (Sec.~\ref{sec:res}).
Finally, we conclude in Sec.~\ref{sec:conclusion}.

\section{SPDC+M nodes}\label{sec:BB}

\begin{figure}[t]
\centering
\includegraphics[width=\linewidth]{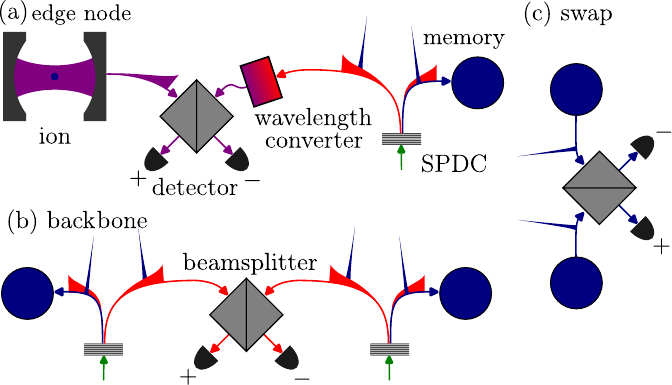}
\caption{\label{fig:modules}
  Sketch of the fundamental steps of the single-click protocol.
  In (a), we show the optical setup of an edge node (EN) to interface an ion with a single rail SPDC+M node.
  In this step, wavelength conversion to match the central frequency combined with the multi-mode nature of the photonic wave function before the detection is used to match the SPDC photon field to the ion.
  (b) The backbone (BB) optical setup to generate long-distance entanglement between remote memories using a central heralding station. 
  In both (a) and (b) a click of detector $+$ or $-$ heralds the generation of entanglement.
  (c) An optical setup analogous to (a) and (b) implements entanglement swapping between memories of EN and BB.
  \added[id=BT]{This figure is adapted from subfigures of the companion article Ref.~\cite{LETT}.}
}
\end{figure}

As discussed in the introduction, we aim to use SPDC+M nodes for their multiplexing capability within the BB, see Fig.~\ref{fig:modules}(b).
These nodes combine a spontaneous parametric down conversion (SPDC) source with a multi-mode memory, e.g., implemented using atomic frequency combs \cite{rakonjac-2021-entan_between_telec_photon_deman,haenni-2025-heral_entan_deman_spin_wave}.
For both the considered protocols, which we denote as \emph{two-single-click} and \emph{double-click protocol}, the BB is operated in a \emph{single-click protocol}.
{We will explain the differences between the protocols in more detail in Sec.~\ref{sec:proto}
and will focus on the SPDC+M nodes and single rail BB operation in this section.
Ideally, the SPDC emits two photons with distinct frequencies, which can be separated into two channels, e.g., by using a chromatic beamsplitter.
One photon is directed toward a quantum memory for storage, while the other is routed to interfere with a photon from another node (such as ions or another SPDC+\replaced[id=BT]{M}{memory} system) and is ultimately detected.
A detector click then serves as a herald for successful entanglement generation between the two nodes.
Additionally, the temporal information associated with the click can be used to initiate storage of the optical excitations in a long-lived spin-wave \cite{rakonjac-2021-entan_between_telec_photon_deman} or to post-select the \emph{time-bin} stored in the memory using on-demand readout \cite{teller-2025-solid_state_tempor_multip_quant}.
Restricting the photon storage to such a time-bin helps suppress noise by discarding unwanted photons, e.g., cases where the other photon of the entangled pair was lost during transmission to the detector.

To model this process, it is convenient (although not strictly necessary) to divide the light emitted by the SPDC into time-bins \replaced[id=BT]{of duration $T_b$ ($T_{\text{BB}}$ within the BB), which can be understood as the time window that contains the state that will be stored in the memory}{, which we chose to match the duration of an acceptance window of the memory $T_b$ (or $T_{\text{BB}}$ within the BB)}.
Within each time-bin, we then describe the state by \cite{hellebek-2024-charac_multim_natur_singl_photon,LETT}
\begin{align}
  \label{eq:ini-spdc}
& \ket{\Psi_b}  = \Big[ \beta_0 + \beta_1 \int_{\mathbb{R}^2} dt dt' \mu(t) F(t,t') b^{\dag}(t) c^{\dag}(t') \\
& \ + \beta_2 \int_{\mathbb{R}^4} d\vec{t} \mu(t_1) \mu(t_2) G(\vec{t}) b^{\dag}(t_1) b^{\dag}(t_2) c^{\dag}(t_3) c^{\dag}(t_4) \Big] \ket{\emptyset} , \notag
\end{align}
with the photonic creation operators \(b^{\dag}, c^{\dag}\) satisfying $\comm{k^{\dag}(t)}{l(t')} = \delta(t-t') \delta_{k,l}$; ($k,l=b,c$).
\added[id=BT]{
We use the notation \(d\vec{t} = \prod_{k=1}^n dt_k\) for the infinitesimal volume in \(n\) dimensions with variables \(t_k\) (\(k=1,\dots,n\)).
}
We will assume that channel \(b\) is connected to the detector used for the initial state generation, and channel \(c\) to the memory and then later used for optical entanglement swapping.
We can express the vacuum and two-pair amplitudes \(\beta_{0}\) and \(\beta_2\) in terms of the single pair amplitude \(\beta_1\).
The temporal modes \(\mu, F, G\) are normalized and encode the temporal shape and correlation of the pairs.
As \(F,G\) depend on multiple times, this models the multi-mode character of the SPDC source (before a heralding photon is detected).

\section{Protocols}\label{sec:proto}

\begin{figure*}[ht]
\centering
\includegraphics[width=\linewidth]{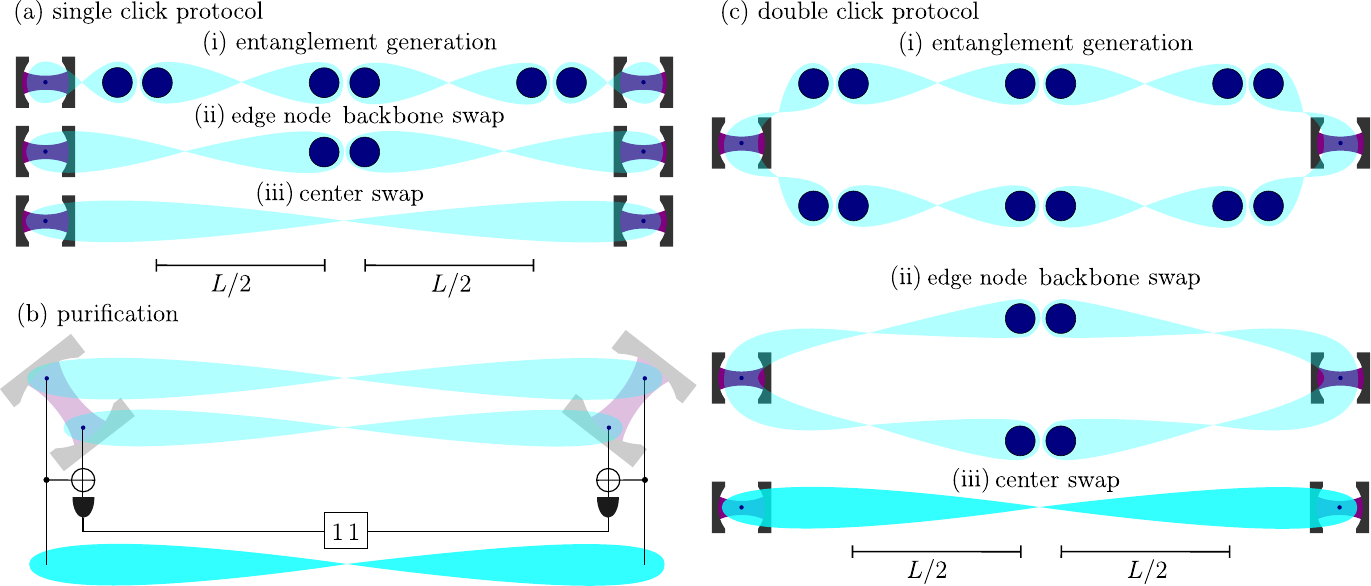}
\caption{\label{fig:protocols}
  Sketch of the entanglement generation protocols, including a multimode repeater node at the center.
  The single-click protocol (a) is visualized using three steps: (i) entangled states are generated between memories in the BB (spanning the long distance \(L/2\)) and between a multimode memory and an ion in the ENs.
  (ii) The range of the ion-photon entanglement is extended to a remote memory using an optical entanglement swap, and finally (iii) another entanglement swap heralds entanglement between the ions.
  (b) Generating entanglement between two ion pairs enables purification using a node-local CNOT gate, followed by read-out of the controlled bits. \added[id=BT]{We term the combination of the single-click protocol (a) with purification (b) the two-single-click protocol.}
  Detecting both in \(\ket{1}\) heralds an entangled state of higher fidelity of the remaining ion pair.
  The double-click protocol (c) uses (i) two clicks to entangle the ions with two rails and then proceeds with similar steps (ii) and (iii) as for the single-click protocol.
  \added[id=BT]{Subfigures (a) and (b) of this figure are adapted from subfigures of the companion article Ref.~\cite{LETT}.}
}
\end{figure*}

Within the edge nodes, we consider two scenarios for the ion operation.
On the one hand, we investigate a single-click approach, where entanglement is generated between the ion and a single memory (rail).
As we will detail below, single-click protocols
require phase stability of the optical paths and
suffer from a growing vacuum component with the length (or entanglement swap depth) \cite{sangouard-2011-quant_repeat_based_atomic_ensem_linear_optic,beukers-2024-remot_entan_protoc_station_qubit}; therefore, we need to generate two ion links that are later distilled into one final pair to suppress the latter issue.
On the other hand, we also develop a double-click protocol to generate a dual rail entangled state with the ion.
This protocol relaxes the phase stability requirement on the ions, but results in worse scaling with intrinsic efficiencies (e.g., memory efficiency) compared to the single-click approach.
The full protocols are illustrated in Fig.~\ref{fig:protocols}.

\subsection{Two-single-click}\label{sec:TSC}

We first discuss the sequential generation of two pairs using a single-click scheme, which are then distilled into one entangled ion pair.
Owing to the single-click nature,
this approach relies on phase stability of the whole setup.
To generate a single pair, the ions of the EN emit a photon such that the state after the emission is described by \cite{LETT,keller-2004-contin_gener_singl_photon_with,barros-2009-deter_singl_photon_sourc_from_singl_ion,tissot-2024-effic_high_fidel_flyin_qubit_shapin,beukers-2024-remot_entan_protoc_station_qubit}
\begin{align}
  \label{eq:ini-atom-SC}
  \ket{\Psi_a} = \left[ \alpha_0 \ket{0} + \alpha_1 \ket{1} \int_{\mathbb{R}} dt \nu(t) a^{\dag}(t) \right] \ket{\emptyset_a},
\end{align}
where we \replaced[id=BT]{create}{have} a single photon \deleted[id=BT]{rail} \(a\added[id=BT]{^{\dag}}\) per ion\replaced[id=BT]{ with }{, and} the single photon emission amplitude \deleted[id=BT]{is} \(\alpha_1\) such that \(\alpha_0 = \sqrt{1 - \abssq{\alpha_1}}\). 
This state is used to generate entanglement between the ion and a memory locally within an EN.
While in parallel entanglement is generated between multi-mode memories in the BB to extend the entanglement over a long distance.

As displayed in Fig.~\ref{fig:protocols}(a), we split the long distance into two, such that an EN and BB state is generated per half.
After successful generation and successful entanglement swapping between the EN and BB memories, the entanglement between the matter-memory-photon entanglement of both halves is extended to the center.
For brevity, we will refer to entanglement swaps as swaps in the following.
Then, a final photonic \deleted[id=BT]{entanglement} swap heralds the successful entanglement of the ions of the ENs.

Alternatively to the setup displayed in the figure, the central repeater node can be omitted.
In the repeater-less setup, we generate entanglement within the two ENs and a BB, followed by optical entanglement swaps between each EN with the BB's memories.
Success of both swaps heralds entanglement between the ions of the ENs.

In practice, we repeat the ion-ion generation for two ions, sharing a single trap, such that we can then use local two-ion gates to purify a single ion pair of higher Bell state fidelity.
The proposed purification consists of local application of a CNOT followed by post-selecting on the target ions in state \(1\), see Fig.~\ref{fig:protocols}(b).
This ensures that the \emph{vacuum} and \emph{two-photon} error are mixed.
As we will detail in the analysis (Sec.~\ref{sec:TSC_ana}), the latter can be suppressed by reducing the emission probability.
By helping in suppressing the former with the latter,
this post-selection leads to a good rate-fidelity tradeoff tailored to this protocol.

\subsection{Double-click}\label{sec:DC}
\begin{figure}[ht]
\centering
\includegraphics[width=\linewidth]{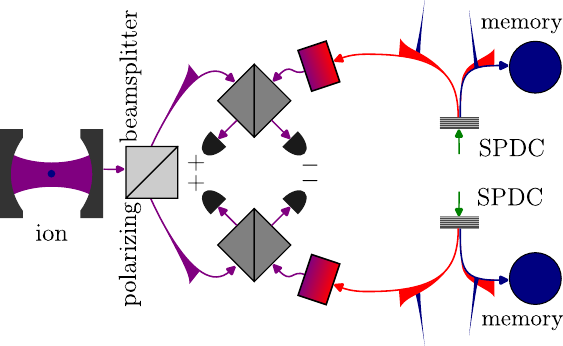}
\caption{\label{fig:DC-EN}
  Sketch of the edge node (EN) setup in the double-click protocol.
  The setup is similar to the single-click EN sketched in Fig.~\ref{fig:modules}(a), but the ion emits photons with two different polarizations entangled with internal states of the ion.
  The polarizations are split using a polarizing beamsplitter (light gray slashed square) and thereby connected to SPDC+M nodes of individual rails.
}
\end{figure}

The second protocol we lay out is a double-click protocol, where the ion emits into two orthogonal modes, which can be coupled to individual \emph{rails} containing SPDC+M nodes, see Fig.~\ref{fig:DC-EN}.
One possibility is to split the modes into different rails using a polarizing beamsplitter if the photons are emitted with different polarizations.
The state of the ions after the full emission is given by
\cite{wilk-2007-singl_atom_singl_photon_quant_inter,stute-2012-tunab_ion_photon_entan_optic_cavit,bock-2018-high_fidel_entan_between_trapp,krutyanskiy-2024-multim_ion_photon_entan_over_kilom}
\begin{align}
  \label{eq:ini-atom-DC}
  \ket{\Psi_a} = \frac{1}{\sqrt{2}} \sum_k  \int_{\mathbb{R}} dt \nu(t) a^{\dag}_k(t) \ket{k} \ket{\emptyset_{a}} ,
\end{align}
with photonic creation operators \(a_k^{\dag}\) within two modes labeled by \(k=0,1\).
We denote the temporal mode of atomic photons by \(\nu\) and the atomic state by \(\ket{k}\) while \(\ket{\emptyset_a}\) represents the shared vacuum over both modes \(a_k\).
By interfacing each of the modes \(a_0\) and \(a_1\) with an SPDC+M node, the matter-photon entanglement is stored in photonic memories, enabling independent rail extension via single-click BB links.
Thus, the protocol retains the beneficial scaling of the single-click protocol with photon losses over long distances \cite{duan-2001-long_distan_quant_commun_with,simon-2007-quant_repeat_with_photon_pair,zhao-2007-robus_creat_entan_between_remot_memor_qubit,sangouard-2011-quant_repeat_based_atomic_ensem_linear_optic,beukers-2024-remot_entan_protoc_station_qubit}.

The full protocol requires generating long distance entanglement in the BB, where we need two successful entangled photon pairs (one per rail).
We consider a scenario where a central SPDC+M repeater is employed, splitting the BB into two parts (see Fig.~\ref{fig:protocols}).
The EN and BB states are prepared in parallel.
After successful entanglement swapping between both EN memories with BB rails on both ends, two swaps at the center (one per rail) herald entanglement between the ions.
Alternatively, without the central repeater node, we connect two hybrid edge nodes with a single BB.
In this case, we try in parallel to generate the EN states as well as the BB state, followed by four optical swaps to generate remote entanglement between the ions.

\section{Protocol analysis}\label{sec:full_ana}

Having sketched the protocols in the previous section, we now turn to the detailed analysis.

\subsection{Detection Model}\label{sec:detection}

Before we investigate the states generated using heralding detection, we discuss a general model of the detection with finite temporal resolution but without detector dead times.
We begin the discussion in the absence of dark counts, but with finite temporal resolution, and without photon number resolution.
Such a detector only discriminates between the presence or absence of any number of photons in the detector within a finite time window.
Including photon losses of the detector in the channel efficiencies,
we model the ideal detection operator corresponding to a single click at time \(t_c\) with temporal resolution \(T\) in one of the ports \(\pm\) as the projection
\begin{align}
  \label{eq:heralding}
  D_{\pm} = D_{\pm}^{\mathcal{T}} (1-D_{\pm}^{\mathbb{R} \setminus \mathcal{T}}) (1-D_{\mp}^{\mathbb{R}}) ,
\end{align}
where the detection operator
\(D_{\pm}^{\mathcal{T}} = \sum_{n=1}^{\infty} \frac{1}{n!} \int_{\mathcal{T}^n} d\vec{t} \left[ \prod_{j=1}^n d_{\pm}^{\dag}(t_j) \right] \proj{\emptyset} \left[ \prod_{j=1}^n d_{\pm}(t_j) \right]\)
projects into having at least one photon in the time interval \(\mathcal{T} = [t_c-T/2,t_c+T/2]\).
Note, that in addition to detection of a photon, as described by the operator $D_{\pm}$, we also project into not having a click outside the time interval and in the other detector, as described by the operators $1-D_{\pm}^{\mathbb{R} \setminus \mathcal{T}}$ and $1-D_{\mp}^{\mathbb{R}}$.
If we consider the action on a single photon state with temporal mode $\nu$, the single photon is thus restricted to the detection interval
\begin{align}
  \label{eq:single-detection}
  D_{\pm}^{\mathcal{T}} \int_{\mathbb{R}} dt \nu(t) d_{\pm}^{\dag}(t) \ket{\emptyset}
  = \int_{\mathcal{T}} dt \nu(t) d_{\pm}^{\dag}(t) \ket{\emptyset} ,
\end{align}
for a two-photon state with temporal modes \(\nu,\mu\) both photons are restricted to the interval
\begin{align}
  \label{eq:double-detection}
  & D_{\pm}^{\mathcal{T}} \int_{\mathbb{R}^2} dt dt' \nu(t) \mu(t') d_{\pm}^{\dag}(t) d_{\pm}^{\dag}(t') \ket{\emptyset} \notag \\
  & = \int_{\mathcal{T}^2} dt dt' \nu(t) \mu(t') d_{\pm}^{\dag}(t) d_{\pm}^{\dag}(t') \ket{\emptyset} .
\end{align}
For a detector with good temporal resolution (i.e., much shorter than the pulse durations of \(\nu,\mu\)), the two photon detection component is suppressed compared to the single photon component, i.e.,
\({T^2 \abssq{\mu(t_c) \nu(t_c)}}/{T \abssq{\nu(t_c)}} = T \abssq{\mu(t_c)} \ll 1\).
Since we consider the emission probability for multiple photons to be (much) smaller than the single photon component, we see that for a detector with good temporal resolution, the single photon components in our model will dominate the two or more photon components, and thus the detector can be treated as approximately number resolving.
\added[id=BT]{
  This regime corresponds to photon pulses much longer than the temporal resolution of the detector.
  We note that state of the art superconducting nanowire single-photon detectors have \added[id=SM]{temporal resolutions on the order of \deleted[id=BT]{tens of} picoseconds~\cite{esmaeil2021superconducting}}, sufficient even for the short memory photons. 
  However, to minimize the dead-time these require mitigation strategies like multi-pixel setups \cite{gruenenfelder-2023-fast_singl_photon_detec_real}.
}
Due to optical losses, however, the detected single photon component can still have contributions from multi-photon parts of the initial state.
In the following, we will assume detectors with sufficient temporal resolution (and no dead time) and thus simplify the calculations by (effectively) treating the detectors as number-resolving.
Additionally, we account for realistic detectors by including the losses in the channel efficiency and adding dark counts as falsely detecting a click while the corresponding channel is in the vacuum state.

\subsection{Backbone state}\label{sec:BB-ana}

Within the BB, we propose the use of a single-click protocol \cite{sangouard-2011-quant_repeat_based_atomic_ensem_linear_optic,beukers-2024-remot_entan_protoc_station_qubit} for long distance entanglement generation and swapping.
This can again be described by a state of the form of Eq.~\eqref{eq:ini-spdc},
but to emphasize that the temporal dependence and amplitudes can be chosen differently in EN and BB, we rename the amplitudes \(\beta \to \gamma\).
Furthermore, we consider a regime where the SPDC emits at a high rate in the BB; therefore, we need to account for an error that can occur if an additional correlated pair is emitted.
For simplicity, we take a single-mode per channel picture (within each time-bin)
\(\mu\to\chi\), \(F(t,t') \to f(t')\), and \(G(\vec{t}) = f(t_3) f(t_4) / 2\),
where the two-pair contribution is described by the product state of the two photon Fock state for both the detection and memory channels
and the probability of emitting two photon pairs is given by \(\abssq{\gamma_2} = \abs{\gamma_1}^4\).
We treat this as an assumption here, but a recent work \cite{hellebek-2024-charac_multim_natur_singl_photon} showed that by appropriately choosing the effective number of modes, the memory photons follow the single-mode statistics upon heralding.

We now turn to the state generated within the BB links by
interfering the photons emitted within a central station using a 50:50 beamsplitter terminating in photon detectors, see Fig.~\ref{fig:modules}(b).
We model the optical channels from left \((1)\) and right \((2)\) to the detection channels $d_{\pm}$ as
\begin{align}
    \label{eq:BB_optics}
b_k \rightarrow \frac{\sqrt{\eta_{\text{BB}}}}{\sqrt{2}} \left[ d_+ + (-1)^k d_- \right] + \sqrt{1-\eta_{\text{BB}}} b_{L,k} .
\end{align}
We show in Appendix~\ref{app:loss} that a combined loss channel \(b_{L,k}\) per node is sufficient if the optical setup is symmetric after the BS.
We divide the BB efficiency $\eta_{\text{BB}}$ in Eq.~\eqref{eq:BB_optics}
into a intrinsic efficiency $\eta_{\text{BB},I}$ and fiber losses quantified by the fiber attenuation length $L_{\text{att}}$, such that
\begin{align}
    \label{eq:eta_BB}
    \eta_{\text{BB}} = \eta_I e^{- L / (n L_{\text{att}})} .
\end{align}
The entire distance $L$ is divided into \(n = 4 (2)\) segments with (without) a repeater.
Assuming a fast (effectively photon number resolving) detector, we find the
click-probability up to second order to be
\begin{align}
  \label{eq:PpmBB}
  P_{\pm}^{\text{BB}} = T \abssq{\chi(t_c)} \eta_{\text{BB}} \abssq{\gamma_1} \left[ 1 + \left(2 - 3 \eta_{\text{BB}}\right) \abssq{\gamma_1} \right] + p_d ,
\end{align}
which heralds a state described by a density matrix
\begin{widetext}
\begin{align}
  \label{eq:RHO_BB}
  \rho_{\pm}^{\text{BB}}
\approx B_0 \proj{\emptyset} + B_1 \proj{\Psi_{\pm}^{\text{BB}}}
 + B_1' \proj{\Psi_{\mp}^{\text{BB}}} 
  + B_2
   \sum_{k=0,2} (\ket{1,1} \pm \sqrt{2} \ket{k,2-k})(\bra{1,1} \pm \sqrt{2} \bra{k,2-k}).
\end{align}
In this description, $B_0$ denotes the probability for the vacuum state in both memories, $6B_2$ the probability to have two photons shared between the memories.
The probability to have a single excitation shared between the memories $B_1 + B_1'$, is made up of the probability $B_1$ to be in the correct Bell state \(\sqrt{2} \ket{\Psi_{\pm}^\text{BB}} = \ket{0,1} \pm \ket{1,0}\) and the probability $B_1'$ to be in the orthogonal one. 
  In the above equations, we assume that the photons traveling from either side to the detector acquire the same phase such that the phases cancel in the off-diagonal elements of the density matrix.
  If however, there is an an unknown phase difference $\phi$ between the left and right fiber channels, the part of the state $\propto B_1$ changes to $\ket{0,1} \pm e^{- i \phi} \ket{1,0}$, reducing the overlap with the ideal Bell state $\ket{\Psi_{\pm}^{\text{BB}}}$.
  This underlines the importance of phase stability for the BB (when employing a single-click protocol).
The normalized elements are approximately given by
\begin{align}
  \label{eq:melBB1}
  B_1 & = \eta_m [1 + (2 - 5\eta_m)(1-\eta_{\text{BB}})\abssq{\gamma_1} - \frac{p_d}{T \abssq{\chi(t_c)} \eta_{\text{BB}} \abssq{\gamma_1}}],
  & B_1' = \abssq{\gamma_1} (1 - \eta_{\text{BB}}) \eta_m(1-\eta_m), \\
  \label{eq:melBB02}
  B_0 & = ( 1-\eta_m ) [1 - 3\eta_m(1-\eta_{\text{BB}})\abssq{\gamma_1}] + \frac{\eta_m p_d}{T \abssq{\chi(t_c)} \eta_{\text{BB}} \abssq{\gamma_1}} ,
  & B_2 = \frac{1}{2} \abssq{\gamma_1} (1 - \eta_{\text{BB}}) \eta_m^2.
\end{align}
\end{widetext}
Here, the the backbone efficiency \(\eta_{\text{BB}}\) and memory efficiency \(\eta_m\) include all losses from the emission to a click event, and \(p_d\) corresponds to the dark-count probability within the detector resolution.
In the limit $p_d = 0$, $\eta_m = 1$, and $\abssq{\gamma_1} \to 0$, we have $B_1 \to 1$ and all other $B_k$'s vanish, i.e., a perfect Bell pair shared between the memories.
Further details of the calculation are provided in Appendix \ref{app:BB}.
The last term of $B_1$ and $B_0$ in Eq.~\eqref{eq:melBB1} and \eqref{eq:melBB02} have a click-time dependent dark-count contribution,
for simpli\replaced[id=BT]{city}{fy} we take the average of this into account by using \({1}/{\abssq{\chi(t_c)}} = T_{\text{BB}}\) for all \(t_c\).

The multimode nature of the SPDC+M nodes naturally enables multiplexing which we quantify by the multiplexing capacity \(N_{\text{BB}}\).
The multiplexing capacity can include a combination of, e.g., spatial, polarization, as well as temporal modes, 
and quantifies the effective number of attempts within the communication time enabled by the additional modes.
The success probability of generating an entangled state within a single attempt is
\begin{align}
  \label{eq:PBB}
\mathbf{P}_{\text{BB}} & = \int_{\mathbb{T}} dt_c {(P_+^{BB} + P_-^{BB})}/{T} \\
   & = 2 \eta_{\text{BB}} \abssq{\gamma_1} \left[ 1 + (2-3\eta_{\text{BB}}) \abssq{\gamma_1} \right] + 2 T_{\text{BB}} \frac{p_d}{T} , \notag
\end{align}
which is enhanced by the multiplexing to ${1 - (1-\mathbf{P}_{\text{BB}})^{N_{\text{BB}}}} \approx N_\text{BB} \mathbf{P}_\text{BB}$ (where the approximation holds in the limit of low success probability).
In Eq.~\eqref{eq:PBB} we integrate over the temporal support of the time-bin \(\mathbb{T}\) of duration $T_{\text{BB}}$.

This analysis of the BB is central to the analysis of both the proposed protocols, because the BB is a key ingredient to achieve high rate entanglement generation. 
Having a model to describe the success probability and density matrix of the memory-memory state enables us to study the state after swapping with the ENs in the following.

\subsection{Two-single-click protocol analysis}\label{sec:TSC_ana}

\subsubsection{Edge node state generation}\label{sec:SC_ENgen}

Having detailed the entanglement generation step within the BB shared between both proposed protocols, it remains to study the generation step in the ENs before we can turn to calculating the state after the optical entanglement swaps between BB and EN memories.
Here we begin with the \deleted[id=BT]{(two)} single-click approach within the ENs, where the initial state of the ions is described by Eq.~\eqref{eq:ini-atom-SC}.
Following the idea outlined in Ref.~\cite{LETT}, we match the weakly driven SPDC source to the ions before heralding.
Within this approach, the matching between the broadband SPDC and the narrow band ion is achieved by modulating the drive
of the SPDC such that the output flux matches the narrowband photon flux of the ions. 
The  heralding click renders the broadband SPDC memory photon into a temporally narrow photon stored in the multimode memory (with a finite efficiency included in $\eta_m$).
As noted in Ref.~\cite{LETT}, we apply wavelength conversion to the SPDC branch to match the ions, as the temporal-post selection of the multimode memory can partially suppress multi-photon losses in the SPDC channel.
Note that due to the weak driving, it is unlikely that emission of two pairs happens simultaneously within the SPDCs \cite{LETT}.
Therefore, we take the EN SPDCs to emit uncorrelated pairs, i.e., for the \(b\) channel we have \(g_2(t,t') \approx 1\).
The SPDC+M state [Eq.~\eqref{eq:ini-spdc}] with uncorrelated pairs satisfies \(2 \abssq{\beta_2} \approx \abs{\beta_1}^4\)
and \(G(\vec{t}) = \delta_1 F(t_1,t_3) F(t_2,t_4) + \delta_2 (t_1 \leftrightarrow t_2)\) with \(2 \abssq{\delta_1 + \delta_2} \approx 1\) \cite{LETT}.

Applying the detection model (good temporal resolution), we calculate the state of the EN ion and memory
\begin{widetext}
\begin{align}
  \label{eq:click_state_EN}
  D_{\pm} \ket{\Psi_a} \ket{\Psi_b} \approx
 \int_{\mathcal{T}} dt d_{\pm}^{\dag}(t) \Big\{
& \alpha_1 \left[ \beta_0 + \beta_1 \sqrt{1-\eta'} \int_{\mathbb{R}^2} dt'dt'' \mu(t') {b}_L^{\dag}(t') F(t',t'') c(t') \right] \ket{1} \sqrt{\frac{\eta}{2}} \nu(t) \\
& \pm \alpha_0 \left[ \beta_1 + \beta_1^2 \sqrt{1-\eta'} \int_{\mathbb{R}^2} dt'dt'' \mu(t') {b}_L^{\dag}(t') F(t',t'') c^{\dag}(t') \right] \ket{0} \sqrt{\frac{\eta'}{2}} \mu(t) \int_{\mathbb{R}} d\tau F(t,\tau) c^{\dag}(\tau) \notag \\
& + \alpha_1 \beta_1 \ket{1} \int_{\mathbb{R}} dt' \nu(t') \sqrt{1-\eta} {a}_L^{\dag}(t) \sqrt{\frac{\eta'}{2}} \mu(t) \int_{\mathbb{R}}d\tau F(t,\tau) c^{\dag}(\tau) \Big\} \ket{\emptyset} , \notag
\end{align}
which still includes the (virtual) loss channels \({a}_L\) and \({b}_L\).
In this expression \(\abssq{\alpha_1}\) and \(\abssq{\beta_1}\) correspond to the ion and SPDC photon emission probability, the temporal modes are determined by \(\nu,\mu,\) and \(F\), and \(\ket{0}\) and \(\ket{1}\) are states of the ion.
The transmission of photons created by \(a^{\dag}\) and \(b^{\dag}\) up to their detection is determined by the efficiencies \(\eta\) and \(\eta'\), respectively.
Accounting for memory losses is possible using \(c = \sqrt{\eta_m} c_m + \sqrt{1 - \eta_m} {c}_L\), with \(\eta_m\) 
the efficiency for detecting photons sent to be stored in the memory, including all losses in between the emission and the detector (within the optical swaps).
We model dark counts as observing a click while the corresponding channel is in vacuum.
Tracing out the loss and detection channels and accounting for dark counts, yields the state after the detection of a click in channel \(\pm\) at time \(t_c\) as
\begin{align}
  \label{eq:RHO_EN}
  \rho_{\pm}^{\text{EN}} 
                             \approx{} \left( A_0 \proj{0} + A_1' \proj{1} \right) \proj{\emptyset} 
  & + A_1 \proj{\varphi_{\pm}} + A_2 \proj{1} \proj{t_c} . 
\end{align}
See Appendix \ref{app:EN} for additional details.
\added[id=BT]{
In the spirit of the temporal filtering and the uncorrelated emission of the SPDC source in the weak driving regime, we treated the non-conditioned memory photon as an orthogonal state that is also traced out.
This is justifies by it being highly unlikely to have two pairs at compatible times in the weak driving limit.
}
Here, the memory state filtered based on the click time \(t_c\) is \(\ket{t_c} = \int_{\mathbb{R}} dt F(t_c,t) c^{\dag} \ket{\emptyset}\)
  and the desired entangled state is \(\ket{\varphi_{\pm}} = \cos\theta \ket{1} \ket{\emptyset} \pm \sin\theta \ket{0} \ket{t_c}\).
  In these expressions, we also introduced the mixing angle of the target state which up to second order is \(\tan^2\theta \approx \frac{\abssq{\mu(t_c)}}{\abssq{\nu(t_c)}} \frac{\eta' \eta_m}{\eta} \frac{\abssq{\beta_1}}{\abssq{\alpha_1}} \left[ 1 - \abssq{\alpha_1} + \abssq{\beta_1} \right]\).
  The angle $\theta$ becomes click-time independent for all time-bins satisfying \(\mu(t) = q \nu(t)\) \added[id=BT]{on the support of the time-bin} (where $q$ is a proportionality constant shared between the time-bins).
  \added[id=BT]{Labeling the time-bins with an index $i$ the combined support ideally spans the full support of the ion temporal mode, i.e., $\sum_{i=1}^N \mu_i(t) = q \nu(t)$.}
The corresponding entries of the density matrix are
\begin{align}
  \label{eq:melEN_O1-first}
  A_1 & = \eta_m \frac{1 + \tan^2\theta}{\eta_m + \tan^2\theta} \left[1 - \frac{\tan^2\theta}{\eta_m + \tan^2\theta} (1-\eta) \abssq{\alpha_1} - \frac{2 \eta_m}{\eta_m + \tan^2\theta} \frac{p_d}{T \abssq{\nu(t_c)} \eta \abssq{\alpha_1}}\right], \\
  \label{eq:melEN_O1-vac}
  A_0 & = (1-\eta_m)  \frac{\tan^2\theta}{\eta_m + \tan^2\theta} \left[1 - \frac{\tan^2\theta}{\eta_m + \tan^2\theta} (1-\eta) \abssq{\alpha_1} + \left( \frac{1}{(1-\eta_m)\tan^2 \theta} - \frac{1}{\eta_m + \tan^2\theta} \right) \frac{2 \eta_m p_d}{T \abssq{\nu(t_c)} \eta \abssq{\alpha_1}} \right], \\
  \label{eq:melEN_O1-last}
A_2 & = \eta_m (1-\eta) \abssq{\alpha_1} \frac{\tan^2\theta}{\eta_m + \tan^2\theta},
      \qquad A_1' = (1-\eta_m) (1-\eta) \abssq{\alpha_1} \frac{\tan^2\theta}{\eta_m + \tan^2\theta} .
\end{align}
Analogous to the BB calculation, we will simplify these by taking the average contribution of dark counts into account by substituting \({\abssq{\nu(t_c)}} = 1/T_a\) (with the ion emission duration $T_a$) for any click time.
Considering the limit where $p_d=0$, $\eta_m=1$, and $\abssq{\alpha_1}, \abssq{\beta_1} \to 0$ (arbitrarily lowering the emission probability while keeping $\tan^2 \theta$ finite), we see that $A_1 \to 1$ while the vacuum component vanishes $A_0 = 0$.
Similarly the two photon component vanishes $A_2 \to 0$ and also $A_1' = 0$.
This corresponds to a perfect Bell state for $\ket{\psi_{\pm}}$ if $\tan^2\theta = 1$, i.e., if the photon flux from both sides are properly matched.
However, we will not fix $\theta$ in the following, because an asymmetry in this state can be corrected by the symmetry around the center of the BB.
We stress that the phase stability arguments for the BB in Sec.~\ref{sec:BB} also apply here, and would introduce a phase to the desired entangled state ($\propto A_1$) \(\cos\theta \ket{1} \ket{\emptyset} \pm e^{- i \phi} \sin\theta \ket{0} \ket{t_c}\).
This underlines the need for phase stability in the EN of the two-single-click protocol.

To calculate the full success probability, we not only need the click probability within a single time-bin of the SPDC source, but also need to account for the fact that only one of the time-bins matching the ion leads to a click, i.e., the remaining time-bins are heralded to be in vacuum.
Therefore, we account for the full initial state \(\ket{\Psi_a} \otimes \prod_{i=1}^N \ket{\Psi_b^i}\) including all \(N\) time-bins;
A single click then results in \(D_{\pm} \ket{\Psi_a} \otimes \prod_i \ket{\Psi_b^i} = D_{\pm} \ket{\Psi_a} \otimes \ket{\Psi_b^k} \prod_{i \ne k} \bra{\emptyset_+} \bra{\emptyset_-} \ket{\Psi_b^i}\), where we have used that only one of the time-bins can lead to a click around time \(t_c\).
Up to first order the probability for a single time-bin being in vacuum is \(p = \abssq{\bra{\emptyset_+} \bra{\emptyset_-} \ket{\Psi_b^i}} \approx 1 - \eta' \abssq{\beta_1}\) such that to first order the probability for \(N-1\) time-bins being in vacuum is \(p^{N-1} \approx 1 - (N-1) \eta' \abssq{\beta_1} \approx 1 - (N-1) \frac{\tan^2\theta}{\eta_m} \frac{\abssq{\nu(t_c)}}{\abssq{\mu(t_c)}} \eta \abssq{\alpha_1}\).
Combined, we find the probability of having a single click over the duration of the ion emission up to second order to be the product
\begin{align}
  \label{eq:click-probability-including-all-time-bins}
P_{\pm} p^{N-1} \approx \frac{\eta}{2} T \abssq{\nu(t_c)} \abssq{\alpha_1} \left\{ \left( 1 + \frac{\tan^2\theta}{\eta_m} \right) \left[ 1 - \eta \frac{\tan^2\theta}{\eta_m} N \frac{\abssq{\nu(t_c)}}{\abssq{\mu(t_c)}} \abssq{\alpha_1} \right] + \frac{\tan^2\theta}{\eta_m} (1-\eta) \abssq{\alpha_1} \right\} + p_d .
\end{align}
Integrating over the entire time interval of the ion emission $\mathbb{T}_I$ with duration $T_a$ and summing over the possible detector clicks we find the success probability
\begin{align}
  \label{eq:PEN}
  \mathbf{P}_{\text{EN}} ={}& p^{N-1} \int_{\mathbb{T}_I} dt_c  \frac{P_+ + P_-}{T} 
  \approx \eta \abssq{\alpha_1} \left\{ \left( 1 + \frac{\tan^2\theta}{\eta_m} \right) \left[ 1 - \eta \frac{\tan^2\theta}{\eta_m}
     \abssq{\alpha_1} \right] \right. 
    + \left. \frac{\tan^2\theta}{\eta_m} (1-\eta) \abssq{\alpha_1} \right\} + 2 T_a \frac{p_d}{T} .
\end{align}
For simplicity, we treat the pulses as constant here, but as the dark counts compete with the real photon probability, it might be beneficial to disregard part of the temporal modes to protect the fidelity \cite{hellebek-2024-charac_multim_natur_singl_photon}.
This corresponds to restricting $\mathbb{T}_I$, thereby reducing the success probability.

\subsubsection{Optical entanglement swaps}
Having calculated the single-click EN state in the previous section and BB state in Sec.~\ref{sec:BB}, we now investigate the state after the probabilistic swaps of the memories.
As depicted in Fig.~\ref{fig:modules}(c), the optical swap between the memories is analogous to the interference process within the entanglement generation stage, i.e., photons retrieved from neighboring memories are combined on a 50:50 beam splitter, and conditioned on a single detection event, the entanglement is transferred to the outer nodes.
In this analysis, we again consider effectively photon number-resolving detectors.
Additionally, we propose using click time information to align the temporal shapes stored in memory, and therefore assume that the memory photons have identical temporal modes. 
This can be achieved by using on-demand read-out of the memories \cite{rakonjac-2021-entan_between_telec_photon_deman,haenni-2025-heral_entan_deman_spin_wave}, which justifies the simplification to a single mode picture.
After the first swap, we find a state of the form
\begin{align}
  \label{eq:RHO_SWAP1}
  \rho_{S1} ={}& \left( C_0 \proj{0} + C_1' \proj{1} \right) \proj{\emptyset} + C_1 \proj{\varphi_{\pm}}
  + \left( C_1'' \proj{0} + C_2 \proj{1} \right) \proj{1_m} \\
  & + C_2' \left[ (\cos\theta \ket{1} \ket{1_m} \pm \sqrt{2} \sin\theta \ket{0} \ket{2_m}) \cdot \text{H.c.} \right]
    + C_3 \proj{1} \proj{2_m}, \notag
\end{align}
which has a similar form as the initially generated state.
However, \(\pm\) now is the product of the detector labels that have clicked, and the memory state (including within the desired entangled state \(\ket{\varphi_{\pm}}\)) refers to the memory with the distance extended by the BB.
We display the non-normalized matrix elements in Appendix \ref{app:optical-swaps}.
\end{widetext}

Due to this similarity, we can treat the repeater and non-repeater cases together by calculating the second swap between two systems with states of the form in Eq.~\eqref{eq:RHO_SWAP1}.
After the second swap, the resulting state between the two ions takes the form
\begin{align}
  \label{eq:final-state}
  \rho = \left(\pm \alpha \op{0,1}{1,0} + \text{H.c.} \right) + \sum_{k,l=0,1} D_{k,l} \proj{k,l} ,
\end{align}
within our perturbative calculation.
We partition the density matrix in the diagonal entries $D_{k,l}$ and the off-diagonal entry $\alpha$.
The sign of the off-diagonal element (\(\pm\)) corresponds to the product of the labels of the detectors that register a click.
Ideally we prepare the Bell state \(\sqrt{2} \ket{\Psi_{\pm}} = \ket{0,1} \pm \ket{1,0}\)
which corresponds to \(D_{0,1}=D_{1,0}=\alpha=1/2\) and all other elements being zero.
In reality, the state is degraded by various imperfections, including multi-photon contributions, dark counts, and memory losses.
These imperfections cause the density matrix to deviate from the ideal state, which is described by the (non-normalized) elements given
in Appendix \ref{app:optical-swaps}.
In the analysis below, we use the analytic expressions for the non-normalized matrix elements of the post-swap states to numerically evaluate the success probability and normalized matrix elements used in further analysis of the protocol.

\subsubsection{Communication Rate}

After calculating the final ion-ion state, as well as the success probabilities of the different steps, we now use these to calculate the duration to prepare the ion-ion state.
We follow the approach of Ref. \cite{avis-2024-asymm_node_placem_fiber_based_quant_networ} to account for parallelization of different operations within the average time it takes to create a link.
Since the initial generation of the fundamental links includes the long distance entanglement generation and the emission of the ions,
we assume that these are the slowest processes.
The duration with a swap at the center then yields
\begin{align}
  \label{eq:duration-center-swap}
  T_{\text{SL}} = \frac{3}{2} \frac{1}{\mathbf{P}_{S2}} \frac{1}{\mathbf{P}_{S1}} \frac{1}{R_{\text{BB}} + R_{\text{EN}}} \left( 1 + \frac{R_{\text{BB}}}{R_{\text{EN}}} + \frac{R_{\text{EN}}}{R_{\text{BB}}} \right) ,
\end{align}
with the entanglement generation rate of the backbone \(R_{\text{BB}} = O_{\text{BB}} N_{\text{BB}} \mathbf{P}_{\text{BB}} / \tau_{\text{BB}}\) and edge node \(R_{\text{EN}} = O_{\text{EN}} \mathbf{P}_{\text{EN}} / \tau_{\text{EN}}\).
In addition to the trial durations $\tau_k$ ($k=\text{BB},\text{EN}$) and generation success probabilities $\mathbf{P}_k$ [see Eqs.~\eqref{eq:PBB} and \eqref{eq:PEN}], these rates depend on the duty cycles $O_k$ for $k=\text{EN}, \text{BB}$, and the BB rate is enhanced by the multiplexing capacity $N_{\text{BB}}$.
We take the typical time-scale for the ENs \(\tau_{\text{EN}}\) to be dominated by the ion-emission $T_a$
and for the BBs \(\tau_{\text{BB}}\) to be dominated by the quantum and classical (back) communication.
We estimate the communication time by twice the light propagation time from the nodes to the heralding station.
The swap probabilities \(\mathbf{P}_{S1}\) and \(\mathbf{P}_{S2}\) both occur only once, because preparation of the extended BB-EN state including the swap ($S1$) happens in parallel in both hal\replaced[id=BT]{v}{f}es; 
These swaps are optically heralded analogous to the generation step (see Fig.~\ref{fig:modules}) leading to the finite success probabilities which are calculated from the non-normalized matrix elements in Appendix \ref{app:optical-swaps}.
The rate-dependent part models the (potentially asymmetric) parallelization of the EN and BB links.
It can be understood as having to wait on average $1/(R_{\text{BB}} + R_{\text{EN}})$ for either the EN or BB to finish. With probability $R_k/(R_{\text{BB}} + R_{\text{EN}})$ module $k$ finished first and we have to wait $1/R_{\bar{k}}$ for the remaining part $\bar{k} \ne k$ to also finish.
The established factor \(3/2\) accounts for the parallelized preparation of two (identical) halves \cite{jiang-2007-fast_robus_approac_to_long,coopmans-2022-improv_analy_bound_deliv_times}. 

Similar to the rate with a central BB repeater, we can also analyze the case of only a single backbone link, where we parallelize the generation of three links.
We express the duration as
\begin{align}
  \label{eq:duration-no-center-swap}
  T_{\text{SL}} ={}& \frac{1}{\mathbf{P}_{S2}} \frac{1}{\mathbf{P}_{S1}} \frac{1}{2R_{\text{EN}} + R_{\text{BB}}} \Bigg[ 1 + \frac{R_{\text{BB}}}{R_{\text{EN}}} \frac{1 + 2 \mathbf{P}_{S1}}{2} \notag  \\
    & + \frac{2R_{\text{EN}}}{R_{\text{EN}} + R_{\text{BB}}} \left( 1 + \frac{R_{\text{EN}}}{R_{\text{BB}}} + \frac{\mathbf{P}_{S1} R_{\text{BB}}}{R_{\text{EN}}} \right) \Bigg] ,
\end{align}
where the first two factors account for the finite swap probabilities.
The rate-dependent fraction and the first term in the square brackets are the average duration up to the first link generation (either EN or BB).
The second term accounts for the fact that with probability \(R_{\text{BB}}/(2R_{\text{EN}}+R_{\text{BB}})\) the first link is the backbone such that after waiting on average for \(1/2R_{\text{EN}}\) we can try the first swap.
After the successful first swap, we need to wait another \(1/R_{\text{EN}}\) before attempting the second swap.
The last term corresponds to the case where an EN was prepared first. In this case, we need to wait for \(1/(R_{\text{EN}}+R_{\text{BB}})\) for the next link, which can be another \(\text{EN}\) (or the BB) in which case we need to wait \(1/R_{\text{BB}}\) (\(1/R_{\text{EN}}\))  before attempting both swaps.
  Note, we assume that if the BB link is generated last, the swaps are attempted simultaneously, i.e., if one of the swaps fails, both ENs and the BB need to be re-generated.
In case the second link is the \(\text{BB}\), we can immediately attempt the first swap and upon success, we wait for the second \(\text{EN}\) and then attempt the final swap.

\subsubsection{Purification}

The analysis up to this point focused on the fundamental single-click protocol.
As noted in the beginning, single-click protocols are known to suffer from an effect called vacuum growth \cite{sangouard-2011-quant_repeat_based_atomic_ensem_linear_optic,beukers-2024-remot_entan_protoc_station_qubit}.
This vacuum growth can also be seen in the preceding analysis, where we see that vacuum admixture in the EN $A_0$ [see Eq.~\eqref{eq:melEN_O1-vac}] and BB $B_0$ [see Eq.~\eqref{eq:melBB02}] cannot be suppressed by lowering the emission probability, while memory losses and dark counts contribute to these components.
Terms mixing these contributions and the target state probabilities within the optical swaps [see Appendix \ref{app:optical-swaps}] cannot be suppressed by lowering the emission probability.
In total, we find that the component $D_{0,0}$ cannot be suppressed by lowering the emission probabilities.
In contrast, the \enquote{double-emission} component \(D_{1,1}\) can be significantly suppressed by lowering the emission probability if the dark-count rate is sufficiently low.
Therefore, we extend the single-click protocol by a purification step, as noted in Sec.~\ref{sec:proto} and sketched in Fig.~\ref{fig:protocols}(b).
We propose to generate two fundamental links, followed by performing local CNOTs between the ions at each end node.
After application of the local gates, we post-select on the target ions being in state \(\ket{1}\).
As we will now show, this will perform a purification of the generated entangled state.
In essence, the purification works because the CNOT and post-selection perform a parity measurement. To lowest order, this excludes the admixture of the desired state $\ket{\Psi_{\pm}}$  with the undesired states $\ket{0,0}$ and $\ket{1,1}$, although a combination of the latter two states can still result in errors; see Eq.~\eqref{eq:non-norm-pur}.

Starting from two entangled ions
\begin{widetext}
\begin{align}
\label{eq:two_single_rails}
  \rho \otimes \tilde{\rho} = & \left( \alpha \ket{0,1'} \bra{1,0'} + \text{H.c.} \right) \otimes \left( \tilde{\alpha} \ket{0,1'} \bra{1,0'} + \text{H.c.} \right)
    + \sum_{k,l=0,1} D_{k,l} \proj{k,l'} \otimes \sum_{\tilde{k},\tilde{l}=0,1} \tilde{D}_{\tilde{k},\tilde{l}} \proj{\tilde{k},\tilde{l}'} \\
  & + \left( \alpha \ket{0,1'} \bra{1,0'} + \text{H.c.} \right) \otimes \sum_{\tilde{k},\tilde{l}=0,1} \tilde{D}_{\tilde{k},\tilde{l}} \proj{\tilde{k},\tilde{l}'}
     + \sum_{k,l=0,1} D_{k,l} \proj{k,l'} \otimes \left( \tilde{\alpha} \ket{0,1'} \bra{1,0'} + \text{H.c.} \right), \notag
\end{align}
where the first (second) rail is left (right) of the tensor product (additionally marked with tildes), and within the rails we use a prime to emphasize that the second bit is in a remote location (i.e., another ion trap) with regard to the first ion.
Local CNOTs (between the unprimed ions as well as between the primed ions) map the state to
\begin{align}
  \label{eq:CNOT-state}
  U_{\text{CNOT}} U_{\text{CNOT}}' \, \rho \otimes \tilde{\rho} \, U_{\text{CNOT}} U_{\text{CNOT}}'= & \left( \alpha \tilde{\alpha} \ket{0,1'} \bra{1,0'} \otimes \ket{0,0'} \bra{0,0'}
  +  \alpha \tilde{\alpha}^{*} \ket{0,1'} \bra{1,0'} \otimes \ket{1,1'} \bra{1,1'} + \text{H.c.} \right) \\
  & + \sum_{k,l,\tilde{k},\tilde{l}=0,1} D_{k,l} \tilde{D}_{\tilde{k},\tilde{l}} \proj{k,l'} \otimes \proj{\abs{\tilde{k} - k},\abs{\tilde{l}-l}'}
  + \dots , \notag
\end{align}
where \(\dots\) are terms \(\propto \alpha \tilde{D}_{k,l},\, \tilde{\alpha} D_{k,l}\) which cannot contribute to the state after measuring the target ions (as they 
are non-diagonal in the measured registers).
Measuring the target register in \({\ket{1,1'}}\) as required  in our purification protocol, we herald the non-normalized density matrix
\begin{align}
 \tilde{\tr}\left[  (\I \otimes \proj{1,1'}) U_{\text{CNOT}} \tilde{U}_{\text{CNOT}} \, \rho \otimes \tilde{\rho} \, U_{\text{CNOT}} U_{\text{CNOT}}' \right]
 =
  & \left(\alpha \tilde{\alpha}^{*} \ket{0,1'} \bra{1,0'} + \text{H.c.} \right)
    + D_{0,1} \tilde{D}_{1,0} \proj{0,1'}  \notag \\
    + D_{1,0} \tilde{D}_{0,1} \proj{1,0'}
\label{eq:non-norm-pur}
    & + D_{0,0} \tilde{D}_{1,1} \proj{0,0'}
    + D_{1,1} \tilde{D}_{0,0} \proj{1,1'} ,
\end{align}
where the trace is over the target qubits.
\end{widetext}
Because the purified state takes the same form as generated state in a fundamental link Eq.~\eqref{eq:final-state},
we can summarize the above result with
the heralding success probability \(P_P = 2 D_{1,0} D_{0,1} + 2 D_{0,0} D_{1,1}\)
and the map describing the state change
\(\alpha_{\text{pur}} = {\alpha^2}/{P_P}\) and \(D_{k,l}^{\text{pur}} = {D_{k,l} D_{1-k,1-l}}/{P_P}\).
For the purified state both \(D_{0,0}^{\text{pur}}\) and \(D_{1,1}^{\text{pur}}\) can be reduced by lowering the emission probability,
since they both involve the $D_{1,1}$ component before the purification.
Including purification and assuming that
the two fundamental links are generated sequentially,
the complete average duration is
\begin{align}
  \label{eq:duration-full}
  T_{\text{TSC}} = 2 T_{\text{SL}} / P_P.
\end{align}
We stress that while this increases the time due to the finite purification probability \(P_P\) (which for the ideal states \(\ket{\Psi_{\pm}}\) is \(1/2\)), it does not increase the scaling in terms of the memory efficiencies \(\eta_m\) and long-range fiber losses, 
because of the deterministic nature of the CNOT operation of the ions.

\subsubsection{Phase stability}
In the preceding analysis, we assumed the system to be phase stable.
But we also noted in the analysis of the generation steps (Sec.~\ref{sec:BB-ana} and \ref{sec:SC_ENgen}), that if there is a difference in the optical paths, it can lead to a phase imprinted on the coherence of the density matrix.
It is instructive to follow this phase in a simplified picture by discarding two-photon components and viewing this phase as between the states $\ket{0}$ (or $\ket{\emptyset}$) and $\ket{1}$, both within the ions and the memories.
If these phases are random and uncorrelated, they generally do not cancel and instead lead to a random phase contribution on the coherence $\alpha$; thereby degrading the Bell state fidelity of the state in Eq.~\eqref{eq:final-state}.
We also see in this simplified picture that if the system is phase stable over both generation steps prior to purification (i.e., the phases can be considered constant and identical between attempts), the purification cancels those phases resulting in the final Bell state, see Eq.~\eqref{eq:non-norm-pur}.
This underlines the phase stability requirement of the two-single-click protocol, also established in the literature \cite{sangouard-2011-quant_repeat_based_atomic_ensem_linear_optic,beukers-2024-remot_entan_protoc_station_qubit}.

\subsection{Double-click protocol analysis}\label{sec:DC_ana}

A possibility to relax the phase stability requirements of the ions in the ENs is the use of
the double-click protocol,
which also helps in avoiding the effect of vacuum growth.
The analysis follows the same approach as described above \added[id=BT]{for the two-single-click protocol}, with the main differences being that we do not perform the purification step, and that the initial EN generation is with two memories; hence we only focus on the key steps in the following.
A numerical implementation of the full protocol is available in Ref.~\cite{CODE}.
First, conditioning on a single click in both the upper and lower arm linked with the first and second rail (see Fig.~\ref{fig:DC-EN}), the state generated at the edge nodes that includes the dual-rail matter-photon entanglement is
\begin{align}
  \label{eq:EN_DC}
  \rho_{\text{EN}}^{\text{DC}} ={}& (A_0' \proj{0} + A_0 \proj{1}) \proj{\emptyset}
+ a \proj{\varphi_{\pm}} \\
& + \sum_{k=0,1} (A_1' \proj{k} + A_1 \proj{1-k}) \proj{1_k} \notag \\
& + A_2 (\proj{0} + \proj{1}) \proj{1_0} \proj{1_1} , \notag
\end{align}
with \(\sqrt{2} \ket{\varphi_{\pm}} = \ket{0} \ket{1_1} \pm \frac{\nu(t_c') \mu(t_c)}{\nu(t_c) \mu(t_c')} \ket{1} \ket{1_0}\) which is equal to the Bell state \(\ket{0} \ket{1_1} \pm \ket{1} \ket{1_0}\) (where the right Ket's encode a dual-rail qubit) if the temporal modes are matched and the system is phase stable over the duration between the click times.
A way to see the latter is by absorbing random phase fluctuations into the mode functions $\mu$ and $\nu$, such that if the phases fluctuate between the click times $t_c$ and $t_c'$, a (unknown) phase is introduced in $\ket{\replaced[id=BT]{\varphi}{\phi}_{\pm}}$.
However, this also underlines that the relevant time-scale, given by the emission duration of the ion, for which the system needs to be phase stable, is relaxed compared to the two-single-click protocol.
Here, we account for the dual rail setup by writing the photonic state \(\ket{k_l}\) with \(k\) photons in rail \(l\).
The entries are given by
\(a/\eta_m = 1 - N \abssq{\beta_1} \frac{\eta'}{\eta} (1-\eta) + \abssq{\beta_1} (1-\eta') + \frac{2 p_d}{T \abssq{\nu(t_c)} \eta' \abssq{\beta_1} N}\),
\(A_2/\eta_m^2 = N \abssq{\beta_1} \frac{1 - \eta}{2} \frac{\eta'}{\eta}\),
\(A_1'/[(1-\eta_m) \eta_m] = A_2/\eta_m^2\),
\(A_1 = A_1' + \eta_m \frac{\abssq{\beta_1}}{2} (1-\eta')\),
\(A_0 = a/\eta_m \frac{(1-\eta_m)}{2} + \frac{(1-\eta_m)^2}{\eta_m^2} A_2 + \frac{1-\eta_m}{\eta_m} (A_1 - A_1') + \frac{p_d}{T \abssq{\nu(t_c)} \eta' \abssq{\beta_1} N}\), and $A_0' = A_0 \vert_{t_c \to t_c'}$.
Note that we again treat dark counts perturbatively, thus disregarding double dark count events.
In the following, we again assume that the modes are ideally matched and for simplicity take the average effect of dark counts into account by substituting $\abssq{\nu(t_c)} \to 1/T_a$ (for any $t_c$).
Here, both rails contain vacuum for the elements $A_0$ and $A_0'$, while for the elements $a,A_1,A_1'$ one photon is in the rails, and for $A_2$ two photons are within the rails.
If we take the limit $\eta_m=1$, $p_d=0$, and $\beta_1 \to 0$, all elements apart from $a \to 1$ approach $0$.

The success probability for the EN generation step within the double-click protocol is
\begin{widetext}
\begin{align}
  \label{eq:PEN_DC}
  \mathbf{P}_{\text{EN}} 
  \approx
  \abssq{\beta_1} \eta' N [\eta + \eta \abssq{\beta_1} + \eta' N \abssq{\beta_1} - (3N-1) \eta \eta' \abssq{\beta_1}]
  + \eta 2 T_a \frac{p_d}{T} 
  ,
\end{align}
\end{widetext}
where we have integrated both rails over the whole ion pulse duration.
We note that there can be additional effects leading to (partially) distinguishable photons, e.g., observed in Ref.~\cite{krutyanskiy-2023-entan_trapp_ion_qubit_separ_by_meter} for ion-ion entanglement generation where a correlation window was used to improve the indistinguishability.
This reduces the accepted click times of the second photon, and thus also the success probability.

There are many different swaps that need to be done, both for the situation with and without the central repeater.
We will not present the full derivation here, but the results are contained in the numerical implementation \cite{CODE}.
Here we only note that if the dual-rail entangled state is extended by two BB \added[id=BT]{links}, such that both rails are extended in distance, we find a state of the form
\begin{align*}
  & \sum_{k=0,1} \sum_{l,m=0,1,2} C_{k,l,m} \proj{k} \proj{l_0} \proj{m_1} \\
  & \pm c \left( \op{0}{1} \op{0_0}{1_0} \op{1_1}{0_1} + \text{H.c.} \right) \\
  & \pm c_2 \left( \op{0}{1} \op{0_0}{1_0} \op{2_1}{1_1} + \text{H.c.} \right) \\
  & \pm c_2' \left( \op{0}{1} \op{1_0}{2_0} \op{1_1}{0_1} + \text{H.c.} \right) \\
  & \pm c_3 \left( \op{0}{1} \op{1_0}{2_0} \op{2_1}{1_1} + \text{H.c.} \right).
\end{align*}
As is visible from this, these states become increasingly complicated.
Therefore, we rely on the numerical simulation \cite{CODE} to calculate the final fidelity and duration.
In the numerical analysis, we again assume that the memory photons share the same shape.
After all optical swaps the state takes the same form as for the single-click protocol Eq.~\eqref{eq:final-state} and the matrix elements can be calculated numerically.

\subsubsection{Communication Rate}\label{sec:DC-rate}
As for the two-single-click protocol, we follow Ref.~\cite{avis-2024-asymm_node_placem_fiber_based_quant_networ} to account for asymmetric simultaneous entanglement generation in calculating the average duration to create an ion-ion link.
Again, for simplicity, we assume that if a central repeater is employed, the resulting halves are both prepared, and then the final swaps are performed at the center.
Thus, we find the total duration to generate an ion-ion link with a central repeater
\begin{align}
  \label{eq:duration_DC_rep}
  T_{\text{DC}} ={}& \frac{3}{2} \left[ \frac{1}{R_{\text{BB}} + R_{\text{EN}}} \left( 1 + \frac{R_{\text{BB}}}{R_{\text{EN}}} + \frac{R_{\text{EN}}}{R_{\text{BB}}} \right) \frac{1}{\mathbf{P}_{S1}} + \frac{1}{R_{\text{BB}}} \right] \notag \\
  & \times \frac{1}{\mathbf{P}_{S2} \mathbf{P}_{S3} \mathbf{P}_{S4}} .
\end{align}
The entanglement generation rate of the backbone \(R_{\text{BB}} = O_{\text{BB}} N_{\text{BB}} \mathbf{P}_{\text{BB}} / \tau_{\text{BB}}\) and edge node \(R_{\text{EN}} = O_{\text{EN}} \mathbf{P}_{\text{EN}} / \tau_{\text{EN}}\) have the same form as for the single-click protocol, but
\(R_{\text{EN}}\) can deviate, due to different duty cycle $O_k$ ($k=\text{BB},\text{EN}$), probability, and duration of the edge node preparation for the different ion-photon state.
The factor of \({3}/{2}\) \cite{jiang-2007-fast_robus_approac_to_long,coopmans-2022-improv_analy_bound_deliv_times} accounts for the cost of preparing each half in parallel, where in each half two BBs are generated sequentially but in parallel with the EN.
As soon as a BB and the EN are ready, the first swap can be performed, and upon success of the swap, it takes on average \(1/R_{\text{BB}}\) to prepare the second BB in order to perform the second swap.
Finally, upon both halves being successfully prepared, the last two swaps ($\mathbf{P}_{S3}$, $\mathbf{P}_{S4}$) are performed.
In addition to the rate with a repeater, we discuss the communication rate without the central repeater in Appendix \ref{app:double-click-no-repeater-rate}.

\section{Results}\label{sec:res}
\begin{figure*}[ht]
\centering
{
\includegraphics[width=0.7\linewidth]{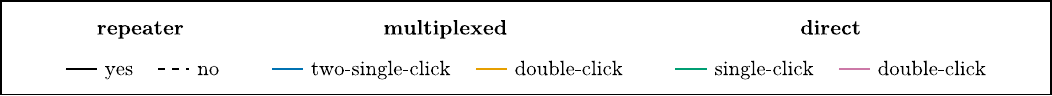} \\
\includegraphics[width=\linewidth]{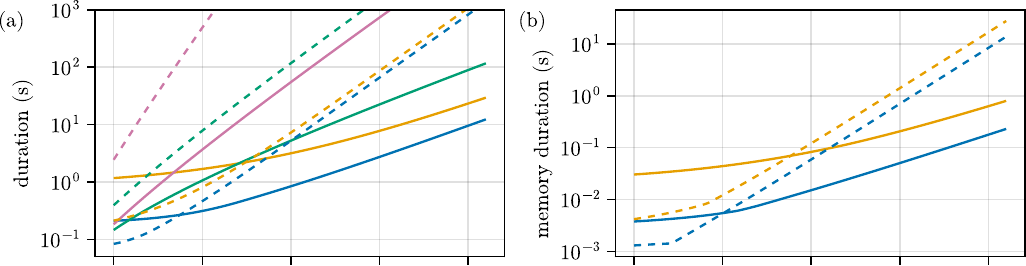} \\
\includegraphics[width=\linewidth]{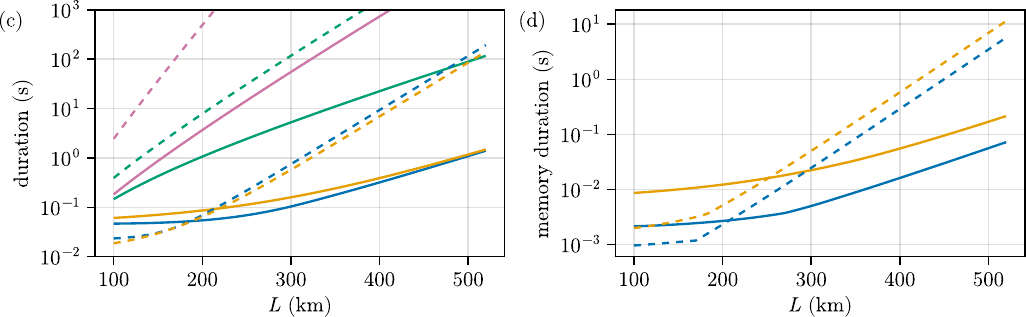}
}
\caption{\label{fig:duration_lower_effs}
\label{fig:duration}
  (a,c) Average preparation duration for remote ion-ion entanglement and (b,d) average worst case storage duration within the multimode memory as functions of ion-ion distance.
  The color encodes the protocol, where blue (orange) is the multiplexed two-single-click (double-click) protocol using the hybrid edge nodes and green (pink) is the direct single-click (double-click) between ions, see legend. 
  Additionally, we compare a repeater-less setup (dashed lines) with the use of a central repeater node (solid lines).
  We take the central repeater to be an SPDC+M node for the multiplexed protocols and an ion repeater for the direct protocols.
  We use the simplified parametrization introduced in the main text with
  a target fidelity \(F_T=0.9\),
  global efficiency \(\eta_I = 0.8\), \added[id=BT]{fiber attenuation of $0.2\,$dB$/$km,}
  edge node duty cycle \(O_{\text{EN}}=0.1\) (to account for cooling and initialization of the ions),
  \(N_{\text{BB}}=1000\) multiplexing channels,
  dark count rate of \(10^{-3}\,\text{s}^{-1}\),
  and an ion-photon duration of \(T_a = 10\,\){\textmu}s,
  which we match with $N=10$ SPDC time-bins.
  Finally we use a memory efficiency \(\eta_m =0.5\) (a,b) or \(\eta_m =0.8\) (c,d).
}
\end{figure*}

Having analyzed the protocols we now turn to evaluating their performance.
For both of the protocols with or with-out the repeater in the center, we find that the final ion-ion state takes the form given in Eq.~\eqref{eq:final-state}.
For the product \(\pm\) of all detector labels that have clicked, we ideally prepare the Bell state \(\sqrt{2} \ket{\Psi_{\pm}} = \ket{0,1} \pm \ket{1,0}\).
The corresponding Bell state fidelity of the state in Eq.~\eqref{eq:final-state} is
\begin{align}
  \label{eq:bell-fidelity}
  F = \braket{ \Psi_{\pm} | \rho | \Psi_{\pm} } = \frac{D_{0,1} + D_{1,0}}{2} + \Re( \alpha ) ,
\end{align}
which is \(1\) if \(D_{0,1} = D_{1,0} = \alpha = 1/2\).

To compare the protocols, we use our numerical implementation (available as Ref.~\cite{CODE}),
where we calculate the average preparation duration by optimizing the emission probabilities to achieve at least a target fidelity \(F_T\).
Note, that for the two-single-click protocol, the ion emission probability is optimized in addition to the SPDC emission probability in the BB and the EN.
In contrast for the double-click protocol, the ion emits a full photon evenly distributed over the two rails [see Eq.~\eqref{eq:ini-atom-DC}] and only the two SPDC emission probabilities are optimized.
Additionally, we compare our protocols using the hybrid nodes to the corresponding duration to generate the ion-ion entanglement directly with a single-click and double-click protocol.
A summary of the derivation for the direct ion-ion generation is given in Appendix~\ref{app:direct}, details of the deterministic entanglement swap using ions as repeaters in Appendix~\ref{app:ion-swaps}\replaced[id=BT]{. In the following, we will investigate results obtained by numerically optimizing the duration to create an ion-ion link of a certain fidelity. We provide }{, and} additional details of the optimization in Appendix \ref{app:opt}.

In Fig.~\ref{fig:duration} 
we simplify the parameter space by only considering two different efficiencies to express the efficiencies of the full model.
We use an intrinsic efficiency \(\eta_I = 0.8\) to account for intrinsic losses.
The efficiencies of the model are then given by \(\eta = \eta_I\) for the ion photon and \(\eta' = \eta_I^2\) \added[id=BT]{for the} SPDC photon \added[id=BT]{being} detected in the generation step, to account for the additional effect of the wavelength conversion only affecting \(\eta'\).
Note, that we proposed to put the wavelength conversion in the SPDC channel, because the increased SPDC emission probability in combination with temporal post selection to reduce the effect of  multi-photon  emissions can suppress part of the effects induced by SPDC photon losses.
The losses in the backbone follow Eq.~\eqref{eq:eta_BB} and we take the intrinsic efficiency \(\eta_{\text{BB},I} = \eta_I\) and fiber loss rate of $0.2\,$dB$/$km which leads to an attenuation length \(L_{\text{att}} \approx 21.7\,\text{km}\).
We take the BB to have a multiplexing capacity of \(N_{\text{BB}}=1000\) and account for a dark-count rate of \(p_d/T = 10^{-3}\,\text{s}^{-1}\).
The ion pulse width of \(T_a = 10\,\){\textmu}s is matched with $N=10$ SPDC time-bins in the simulation corresponding to a photon acceptance interval of $T_b = 1\,${\textmu}s.
We assume that the detection window relevant for the heralding of a memory-memory entanglement swap is a tenth of a time-bin $T\abssq{f(t_c)} = 10 T\abssq{\chi(t_c)} = 10 N T \abssq{\nu(t_c)} = 10 N T / T_a$.
Note that for simplicity we take the pulses as constant over their duration and the ion and SPDC to be ideally matched [see also Eqs.~\eqref{eq:PBB}, \eqref{eq:PEN}, and \eqref{eq:PEN_DC}].
To account for the contrast of the ideally continuous operation of the BB versus the need to reset the ions after every unsuccessful try, we use \(O_{\text{EN}}=0.1\) and \(O_{\text{BB}}=1.0\).
We choose a target fidelity \(F_T = 0.9\),
and show how the different protocols are affected by the memory efficiency
using \(\eta_{m} = 0.5\) in Fig.~\ref{fig:duration}(a,b) and \(\eta_m=0.8\) in Fig.~\ref{fig:duration}(c,d).

We see that already for the lower memory efficiency \(\eta_m=0.5\) [Fig.~\ref{fig:duration}(a,b)] our protocols shows a significant speed-up compared to direct ion-ion entanglement generation.
For the two-single-click
protocol we observe a speed-up of about an order of magnitude compared to the direct ion single-click protocol, and for the double-click protocol about more than two orders of magnitude compared to the direct ion double-click protocol.
If we compare the durations of the multiplexed two-single-click and double-click protocols, we see that the two-single-click protocol is faster.
This is due to a better scaling with the swap success probability since it requires fewer simultaneous successful photon detection events,
and thus also has a better scaling with the memory efficiency.

For a higher memory efficiency [\(\eta_m=0.8\) in Fig.~\ref{fig:duration}(c,d)], we observe that the durations of the multiplexed protocols do not deviate significantly.
For both protocols the higher memory efficiency increases the speed-up of the protocols compared to the non-multiplexed versions.

In Fig.~\ref{fig:duration}, we also display the average worst case ensemble memory storage duration,
which we introduce as a quantity accessible within our model that quantifies the duration photons are stored within the memory.
We calculate it as the maximum over the configurations of the longest average waiting time of a memory, e.g., for the two-single-click protocol with a central repeater it is given by $1/\min(R_{\text{BB}}, R_{\text{EN}}) + T_{\text{half}}$ where $T_{\text{half}} = 2 \mathbf{P}_{S2} T_{\text{SL}}/3$ is the average duration to prepare one half of the repeater, see also the numerical implementation~\cite{CODE}.
This helps in determining the feasibility of a protocol, as realistic memories have a finite lifetime \cite{zhao-2008-millis_quant_memor_scalab_quant_networ}.
If the lifetime is much longer than the storage duration, the constant efficiency we employed in our model is a good approximation and underlines the feasibility of the protocol.
In Fig.~\ref{fig:duration}(b,d) we see that the two-single-click protocol has a lower average memory storage duration for both considered memory efficiencies.
Thus, the two-single-click protocol can be used for shorter lived memories compared to the double-click protocol.
Furthermore both protocols, achieve a lower memory storage duration constraint for the higher memory efficiency.
In all cases there is an additional requirement that ions remain coherent over the duration of the protocol. However, trapped ions have already demonstrated long coherence times, compatible with the multiplexed protocols \cite{wang-2021-singl_ion_qubit_with_estim,drmota-2023-robus_quant_memor_trapp_ion}. We therefore focus on the memory coherence time.

\added[id=BT]{
  In realistic near-term implementations, memory efficiencies will likely vary among different memories and decay within the storage time.   One approach to deal with the decaying memory efficiency is the introduction of a time-out resulting in a non-trivial optimization problem making it hard to analyze in detail \cite{davies-2024-tools_analy_quant_protoc_requir}.
  Based on our analysis, however, we expect that the two-single-click protocol tolerates lower memory decay times compared to the double-click protocol, since it is generally more robust to losses.
  Similarly, we expect the two-single-click protocol to be less affected by non-uniform memory efficiencies, because
 a naive mitigation strategy of (known) non-uniform memory efficiencies is introducing additional loss to make the memories effectively uniform, thus lowering the overall efficiency. Non-uniform memory efficiencies might have more desirable mitigation strategies and it is therefore hard to conclude something in general. 
  A complete investigation on these questions will be highly relevant for near term implementations, but is beyond the scope of this work.
}

\added[id=BT]{
  Another important factor for implementations is phase stabilization.
  Actively stabilizing the fiber length using a classical signal for links between SPDC+M nodes using a single-click protocol has already been demonstrated \cite{lago-rivera-2021-telec_heral_entan_between_multim,haenni-2025-heral_entan_deman_spin_wave} and is necessary for all the protocols under consideration.
  However, the two-single-click protocol requires the phase of the ions to remain stable from photon emission until both ion-ion links are established. 
  While achieving this may be challenging, it can build upon existing techniques for cavity stabilization \cite{stolz-2022-quant_logic_gate_between_two} as well as ion–ion entanglement \cite{liu-2026-long_lived_remot_ion_ion}.
  This requirement is relaxed by the double-click protocol, where it is necessary for the phase to remain stable between the two detection events during the EN \added[id=AS]{entanglement} generation.
  For direct double-click ion-ion entanglement generation the relevant time-frame in a recent experiment \cite{krutyanskiy-2023-entan_trapp_ion_qubit_separ_by_meter} was in the order of $10\mu$s.
  If the phase cannot be stabilized or corrected over the relevant timescale, this will introduce an additional error on the final state which needs to be factored into the total error budget, e.g., by (if possible) targeting a fidelity high enough such that even with the phase error the fidelity requirements are satisfied.
  A simple way to estimate the error is found by investigating a random relative phase $\phi$ on the $\ket{1}$ state of one qubit.
  This affects the off-diagonal element $\alpha$ [see Eq.~\eqref{eq:bell-fidelity}] and reduces the fidelity of an ideal Bell pair to $(1 + \cos \phi ) / 2$.
}

In both the considered examples we see that the hybrid nodes can lead to a speed-up compared to pure ion nodes, due to the multiplexing capacity of the SPDC+M BB.
Furthermore, if phase stability can be achieved and the memory efficiencies are not approaching unity, the single-click protocol is favorable.
On the other hand, since we observe no significant speed-up of the multiplexed \replaced[id=BT]{two-single-click}{double-click} protocol compared to the \replaced[id=BT]{double-click}{two-single-click} protocol for high memory efficiencies,
the relaxed phase stability requirements, makes th\replaced[id=BT]{e double-click}{is} protocol favorable in the domain of high memory efficiency if the memory storage duration allows it.

We note that the hybrid protocols can lead to more resilience regarding dark counts, as was also shown in Ref.~\cite{LETT}.
This can be attributed to a single dark count leading to a parity that is purified in the two-single-click protocol.
Furthermore,
this aligns well with the fact that the hybrid protocol splits the constraint regarding the dark count in two parts one with a low efficiency due to fiber losses combined with a shorter pulse and one with a higher efficiency in combination with \deleted[id=BT]{a} the longer ion pulse \(\abssq{\nu(t_c)} \sim 1/T_a\).
In contrast, the direct ion-ion entanglement generation is limited by both simultaneously.
Here we have considered up to one SPDC+M repeater node within the chain, enabling high-rate entanglement generation over a few hundred kilometer.
As probabilistic entanglement swapping limits the entanglement generation rate for long distances \cite{simon-2007-quant_repeat_with_photon_pair},
we propose using deterministic ion-ion swaps \cite{riebe-2008-deter_entan_swapp_with_ion} to further increase the connectivity and or range of the network.

\section{Conclusion}\label{sec:conclusion}

We have investigated the implementation of a (two) single-click and a double-click protocol using our recently proposed hybrid nodes \cite{LETT}.
The hybrid nodes unite ions, spontaneous parametric down conversion sources, and ensemble memories and solve the mismatch of the photon bandwidth between those systems in the initial matter-photon entanglement generation \cite{LETT}.
Here, we have provided a thorough analysis of the ion-ion entanglement generation that we envision for these hybrid nodes using the two different protocols.
We find that for optimistic near term parameters, the use of hybrid nodes can enable fast remote ion entanglement generation by providing the multiplexing capacity of the SPDC and ensemble memories both for single and double-click protocols.
While the single-click protocol puts less demand on the ensemble memories, it requires phase stability including the ions.
Therefore we conclude that for good memories, the double-click protocol is favorable,
while the single-click is favorable if phase stability can be achieved for the ion out-coupling but the memory has some losses.

The proposed architecture combines the distinct advantages of the high multiplexing capacity of ensemble based memories with the processing capabilities of trapped ions.
We envision that this can be used to construct a quantum internet where processing nodes are connected by high-speed multi-mode quantum memories for long distance communication.
As we have seen, the memory efficiency has a significant influence on the protocols,
therefore it 
will be important to increase the memory efficiency and coherence time 
but also to perform more detailed theoretical investigations of the effect of the memory efficiency, e.g., how time-dependent memory efficiencies affect different protocols.

\begin{acknowledgments}
  We thank T. E. Northup, H. de Riedmatten, S. D. C. Wehner, M. van Hooft, N. Sangouard, P. Cussenot, B. Grivet, S. Grandi, \deleted[id=BT]{and} A. Das\added[id=BT]{, and C. Gustin} for fruitful discussions.
    This work was funded by the European Union's Horizon Europe research and innovation programme under grant agreement No.~101102140 – QIA Phase 1.
  Funded by the European Union.
  Views and opinions expressed are however those of the authors only and do not necessarily reflect those of the European Union or European Commission.
  Neither the European Union nor the granting authority can be held responsible for them.
  BT, ERH and ASS acknowledge the support of Danmarks Grundforskningsfond (DNRF Grant No.~139, Hy-Q Center for Hybrid Quantum Networks).
\end{acknowledgments}

\appendix
\begin{figure}[ht]
\centering
\includegraphics[width=7cm]{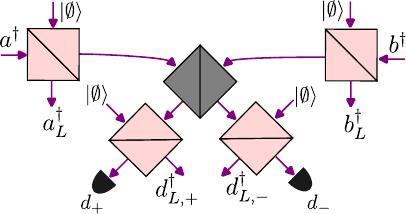}
\caption{\label{fig:loss}Sketch of the optical setup for single-click entanglement generation. Two channels \(a\) and \(b\) are combined using a beamsplitter (gray) whoose output ports terminate in detectors \(d_\pm\). Losses in the channels and detectors are modeled using BS (light pink) with additional vacuum input $\ket{\emptyset}$ and output into loss channels $\cdot_L$.}
\end{figure}
\section{Effective channel losses in beamsplitter combined detection}\label{app:loss}

In this appendix we describe and justify the model we use to account for optical losses in the main text.
We model the optical setup of the entanglement generation including various losses,
where we describe the losses using virtual channels connected to physical channels with beam-splitters, see Fig.~\ref{fig:loss}.
The creation operator of a photon at system A (B) \(a^{\dag}\) (\(b^{\dag}\)) then experiences the transformation
\begin{widetext}
\begin{align}
  \label{eq:optics_general}
  \left. \begin{matrix} a^{\dag} \\ b^{\dag} \end{matrix} \right\} & \to \begin{cases} \sqrt{\eta_a} \\ \sqrt{\eta_b} \end{cases} \left[ \pm \cos\phi \left( \sqrt{\eta_{\pm}} d_{\pm}^{\dag} + \sqrt{1 - \eta_{\pm}} {d}_{L,\pm}^{\dag} \right) + \sin\phi \left( \sqrt{\eta_{\mp}} d_{\mp}^{\dag} + \sqrt{1 - \eta_{\mp}} {d}_{L,\mp}^{\dag} \right) \right] + \begin{cases} \sqrt{1 - \eta_a} {a}_L^{\dag} \\ \sqrt{1 - \eta_b} {b}_L^{\dag} \end{cases} ,
\end{align}
\end{widetext}
where we introduced the annihilation operators of the detector \(d_{\pm}\)
and the branching ratio of the physical BS is determined by the branching angle \(\phi\).
The operators \({\cdot}_L\) denote loss channels,
where
\({d}_{L,\pm}\) combines losses after the BS with losses of the detector and \({a}_L, {b}_L\) include all losses before the BS and potential losses of the BS.
For brevity we introduce the combined loss operators
\begin{align}
  \label{eq:losses}
  \underline{a}^{\dag} & = a^{\dag} - \sqrt{\eta_a \eta_+} \cos\phi d_+^{\dag} - \sqrt{\eta_a \eta_-} \sin\phi d_-^{\dag}, \\
  \underline{b}^{\dag} & = b^{\dag} - \sqrt{\eta_b \eta_+} \sin\phi d_+^{\dag} + \sqrt{\eta_b \eta_-} \cos\phi d_-^{\dag} ,
\end{align}
which satisfy the commutation relations
\begin{align}
  \label{eq:loss_commutation}
  \comm{\underline{a}}{\underline{a}^{\dag}} = 1 - \eta_a \left( \cos^2\phi \eta_+ + \sin^2\phi \eta_- \right) , \\
  \comm{\underline{b}}{\underline{b}^{\dag}} = 1 - \eta_b \left( \sin^2\phi \eta_+ + \cos^2\phi \eta_- \right) , \\
  \comm{\underline{a}}{\underline{b}^{\dag}} = \sqrt{\eta_a \eta_b} \sin\phi \cos\phi \left(  \eta_+ - \eta_- \right) .
\end{align}
These show that if the losses after the BS are equal, i.e., \(\eta_+ = \eta_-\), the two effective loss operators commute \(\comm{\underline{a}}{\underline{b}^{\dag}} = 0\).
We consider this scenario in our analysis of the protocol which furthermore employs a balanced BS so \(\cos\phi = \sin\phi = {1}/{\sqrt{2}}\).
Therefore, we summarize the losses in a single loss channel \({a}_L, {b}_L\) per physical channel \(a,b\).

\begin{widetext}
\section{Additional information on the derivation of the BB state}\label{app:BB}

In the main text we introduced the state heralded within the BB [Eq.~\eqref{eq:RHO_BB}].
In the following, we provide additional details of the derivation.
Based on the SPDC+M model discussed in Secs.~\ref{sec:BB} and \ref{sec:BB-ana}.
Using the SPDC+M initial state [see Eq.~\eqref{eq:ini-spdc}], we can express the un-normalized heralded state including the loss channels and already applying the memory losses
\begin{align}
  \label{eq:click_state_BB}
  D_{\pm} \ket{\Psi_l} \ket{\Psi_r} \approx{} &
\sqrt{\eta_{\text{BB}}} \int_{\mathcal{T}} dt \chi(t) d_{\pm}^{\dag}(t) \Big\{
 \gamma_0 \gamma_1 \frac{c_{f,l}^{\dag} \pm c_{f,r}^{\dag}}{\sqrt{2}} \\
&\qquad\quad + \frac{\sqrt{1 - \eta_{\text{BB}}}}{\sqrt{2}} \int_{\mathbb{R}} dt'' \chi(t'') \Big[
{b}_{L,l}(t'') c_{f,l}^{\dag} (\gamma_0 \gamma_2 c_{f,l}^{\dag} \pm \gamma_1^2 c_{f,r}^{\dag})
+ {b}_{L,r}(t'') (\gamma_1^2 c_{f,l}^{\dag} \pm \gamma_0 \gamma_2 c_{f,r}^{\dag}) c_{f,r}^{\dag}
 \Big] \Big\} \ket{\emptyset} , \notag
\end{align}
with \(c_{f,k} = \int_{\mathbb{R}} dt f(t) \left[ \sqrt{\eta_m} c_{m,k}^{\dag}(t) + \sqrt{1 - \eta_m} {c}_{L,k}^{\dag}(t) \right]\) and \(k=l,r\).
For both channels \(c_{m,k}^{\dag}\), \({c}_{L,k}^{\dag}\), and \({b}_{L,k}^{\dag}\) respectively create a photon in the memory channel, memory loss channel or initial heralding loss channel; \(d_{\pm}^\dag\) creates a photon in detector channel \(\pm\).

As stated in the main text to account for dark counts,
we assume that a vacuum state in the detector channel can falsely lead to a click, such that the post click density matrix becomes
\begin{align}
  \label{eq:dark}
  P_{\pm}\added[id=BT]{^{\text{BB}}} \rho\added[id=BT]{_{\pm}^{\text{BB}}} = \tr_{\text{photon}} \added[id=BT]{(} D_{\pm} \proj{\Psi_l} \proj{\Psi_r} D_{\pm} + p_d \proj{\emptyset} \proj{\Psi_l} \proj{\Psi_r} \proj{\emptyset} \added[id=BT]{)} ,
\end{align}
with the probability to detect a dark count within the detector resolution \(p_d\) (the dark count rate corresponds to \(p_d/T\)).
We treat \(p_d\) to be of comparable order to the \added[id=BT]{(}mixed\added[id=BT]{)} second order of the emission probabilities.
Tracing out the loss and detection channels results in
\begin{align}
  \label{eq:RHO_BB-non-normalized}
  P_{\pm}^{\text{BB}} \rho_{\pm}^{\text{BB}} 
& \approx \eta_{\text{BB}} \abssq{\gamma_1} (1 - \abssq{\gamma_1}) T \abssq{\chi(t_c)} \left[ \eta_m \proj{\Psi_{\pm}^{\text{BB}}} + (1 - \eta_m) \proj{\emptyset} \right]
 + p_d \proj{\emptyset}
\\
& \qquad + \frac{\eta_{\text{BB}}}{2} \abs{\gamma_1}^4  T \abssq{\chi(t_c)} (1 - \eta_{\text{BB}}) \big[
\begin{aligned}[t]
&   \eta_m^2 (\pm \ket{1,1} + \sqrt{2} \ket{2,0})(\pm \bra{1,1} + \sqrt{2} \bra{2,0}) \\
& + \eta_m^2 (\ket{1,1} \pm \sqrt{2} \ket{0,2})(\bra{1,1} \pm \sqrt{2} \bra{2,0}) \\
& + \eta_m (1 - \eta_m) (10 \proj{\Psi_{\pm}^{\text{BB}}} + 2 \proj{\Psi_{\mp}}) \\
& + (1 - \eta_m)^2 6 \proj{\emptyset}
 \big],
\end{aligned} \notag
\end{align}
with \(\sqrt{m!n!} \ket{m,n} = \left[ \int_{\mathbb{R}} dt f(t) c_{l,m}^{\dag}(t) \right]^m \left[ \int_{\mathbb{R}} dt f(t) c_{r,m}^{\dag}(t) \right]^n \ket{\emptyset}\) and \(\sqrt{2} \ket{\Psi_{\pm}^{\text{BB}}} = \ket{1,0} \pm \ket{0,1}\).

The non-normalized elements of the density matrix given in the main text thus are
\begin{align}
  \label{eq:melBBUN1}
  P^{BB}_{\pm} B_0 & = T \abssq{\chi(t_c)} \left[ \eta_{\text{BB}} \abssq{\gamma_1} (1 - \abssq{\gamma_1}) + 6 (1 - \eta_m) \frac{\eta_{\text{BB}}}{2} \abs{\gamma_1}^4 (1 - \eta_{\text{BB}}) \right] (1 - \eta_m) + p_d, \\
P^{BB}_{\pm} B_1 & = T \abssq{\chi(t_c)} \left[ \eta_{\text{BB}} \abssq{\gamma_1} (1 - \abssq{\gamma_1}) + 10 (1 - \eta_m) \frac{\eta_{\text{BB}}}{2} \abs{\gamma_1}^4  (1 - \eta_{\text{BB}}) \right] \eta_m , \\
  P^{BB}_{\pm} B_1' & = T \abssq{\chi(t_c)} \frac{\eta_{\text{BB}}}{2} \abs{\gamma_1}^4  (1 - \eta_{\text{BB}}) \eta_m (1 - \eta_m) 2,
  \qquad P_{\pm}^{\text{BB}} B_2 = T \abssq{\chi(t_c)} \frac{\eta_{\text{BB}}}{2} \abs{\gamma_1}^4  (1 - \eta_{\text{BB}}) \eta_m^2 .
\end{align}

\section{Additional information on the calculation of the EN-BB state (single-click)}\label{app:EN}

In this appendix, we provide further details for the calculation of the EN state and the corresponding success probability.
The key details and results of this are discussed in Sec.~\ref{sec:SC_ENgen} of the main text.
Based on the post-detection state in Eq.~\eqref{eq:click_state_EN},
we calculate the density matrix by tracing out the photonic loss (including \({c}_L\)), \added[id=BT]{the filtered part stored in the memory}, and detection channels and disregard the terms that are suppressed due to the temporal separation between the ion emission duration $T_a$ and SPDC correlation time.
This yields
\begin{align}
  \label{eq:RHO_EN-nonnorm}
  \tr_{\text{photons}} D_{\pm} \proj{\Psi_a} \proj{\Psi_b} D_{\pm}
\approx &
\Bigg\{ \added[id=BT]{\left[ \frac{\eta}{2} T \abssq{\nu(t_c)} \abssq{\alpha_1} \abssq{\beta_0} + \frac{\eta' \eta_m}{2} T \abssq{\mu(t_c)} \abssq{\alpha_0} \abssq{\beta_1} \right]} \proj{\varphi_{\pm}}
\notag \\
& \quad + \frac{\eta' (1-\eta_m)}{2} T \abssq{\mu(t_c)} \abssq{\alpha_0} \abssq{\beta_1} \proj{0} \proj{\emptyset} \Bigg\} \left[ 1 + (1-\eta') \abssq{\beta_1} \right]  \notag \\
& + \frac{\eta' (1-\eta)}{2} T \abssq{\mu(t_c)} \abssq{\alpha_1} \abssq{\beta_1} \proj{1} \left[ \eta_m \proj{t_c} + (1-\eta_m) \proj{\emptyset} \right] .
\end{align}
For brevity of the expressions, we introduce
the state stored in the multi-mode memory conditioned on the detection of a click at time \(t_c\) as \(\ket{t_c} = \int_{\mathbb{R}} dt F(t_c,t) c^{\dag} \ket{\emptyset}\). 
The generated target state is \(\ket{\varphi_{\pm}} = \cos\theta \ket{1} \ket{\emptyset} \pm \sin\theta \ket{0} \ket{t_c}\),
where the mixing angle \(\theta\) of the target state is determined by \(\tan\theta = \frac{\sqrt{\eta' \eta_m} \mu(t_c) \alpha_0 \beta_1}{\sqrt{\eta} \nu(t_c) \alpha_1 \beta_0}\) or approximated to second order
\(\tan^2\theta \approx \frac{\abssq{\mu(t_c)}}{\abssq{\nu(t_c)}} \frac{\eta' \eta_m}{\eta} \frac{\abssq{\beta_1}}{\abssq{\alpha_1}} \left[ 1 - \abssq{\alpha_1} + \abssq{\beta_1} \right]\).

We account for dark counts analogous to the previous appendix,
where we treat \(p_d\) to be of comparable order to the \added[id=BT]{(}mixed\added[id=BT]{)} second order of the emission probabilities \(\abssq{\alpha_1}\) and \(\abssq{\beta_1}\), thus the approximate contribution due to dark counts is \(\tr_{\text{photons}} \bra{\emptyset_+} \bra{\emptyset_-} \proj{\Psi_a} \proj{\Psi_b} \ket{\emptyset_+} \ket{\emptyset_-} \approx p_d \proj{\emptyset_a} \proj{\emptyset_b}\).
With this we find the form of the density matrix given in the main text, with the click probability
\begin{align}
  \label{eq:PpmEN_app}
  P_{\pm} ={}& \left[ 1 + (1-\eta') \abssq{\beta_1} \right] \left[ \frac{\eta}{2} T \abssq{\nu(t_c)} \abssq{\alpha_1} \abssq{\beta_0} + \frac{\eta'}{2} T \abssq{\mu(t_c)} \abssq{\alpha_0} \abssq{\beta_1} \right]
             + \frac{\eta' (1-\eta)}{2} T \abssq{\mu(t_c)} \abssq{\alpha_1} \abssq{\beta_1}
             + p_d ,
\end{align}
as well as the non-normalized matrix elements
\begin{align}
  \label{eq:melEN1}
  P_{\pm} A_1 & = \left[ 1 + (1-\eta') \abssq{\beta_1} \right] \left[ \frac{\eta}{2} T \abssq{\nu(t_c)} \abssq{\alpha_1} \abssq{\beta_0} + \frac{\eta' \eta_m}{2} T \abssq{\mu(t_c)} \abssq{\alpha_0} \abssq{\beta_1} \right], \\
  P_{\pm} A_0 & = \left[ 1 + (1-\eta') \abssq{\beta_1} \right] \frac{\eta' (1-\eta_m)}{2} T \abssq{\mu(t_c)} \abssq{\alpha_0} \abssq{\beta_1} + p_d, \\
  \label{eq:melEN3}
  P_{\pm} A_2 & = \frac{\eta' (1-\eta)}{2} T \abssq{\mu(t_c)} \abssq{\alpha_1} \abssq{\beta_1} \eta_m ,
                \qquad P_{\pm} A_1' = P_{\pm} A_2 \frac{1-\eta_m}{\eta_m} .
\end{align}
The above equations are up to \added[id=BT]{(}mixed\added[id=BT]{)} second order in the emission probabilities \(\abssq{\alpha_1}\) and \(\abssq{\beta_1}\) (treating \(p_d\) as second order itself) and we assumed that the SPDC is weakly driven to match the atomic system, see also Ref.~\cite{LETT}.

To express the normalized density matrix elements up to \deleted[id=BT]{mixed} first order, we treat \(\theta\) as a zeroth order quantity and
express \(T \abssq{\mu(t_c)} \eta' \abssq{\beta_1} \approx \frac{\tan^2\theta}{\eta_m} T \abssq{\nu(t_c)} \eta \abssq{\alpha_1} (1-\abssq{\beta_1}+\abssq{\alpha_1})\)
which we can apply to itself to find
\(T \abssq{\mu(t_c)} \eta' \abssq{\beta_1} \approx \frac{\tan^2\theta}{\eta_m} T \abssq{\nu(t_c)} \eta \abssq{\alpha_1} \left[ 1+\abssq{\alpha_1} - \frac{\tan^2\theta}{\eta_m} \frac{\abssq{\nu(t_c)}}{\abssq{\mu(t_c)}} \frac{\eta}{\eta'} \abssq{\alpha_1} \right]\).
We can then first use this to calculate the click probability up to \added[id=BT]{(}mixed\added[id=BT]{)} second order
\begin{align}
  \label{eq:PpmEN_2}
  P_{\pm} ={}&  \frac{\eta}{2} T \abssq{\nu(t_c)} \abssq{\alpha_1} \left\{ \left( 1 + \frac{\tan^2\theta}{\eta_m} \right) \left[ 1 - \eta \frac{\tan^2\theta}{\eta_m} \frac{\abssq{\nu(t_c)}}{\abssq{\mu(t_c)}} \abssq{\alpha_1} \right] + \frac{\tan^2\theta}{\eta_m} (1-\eta) \abssq{\alpha_1} \right\} + p_d .
\end{align}
Substituting the second order in \(\abssq{\alpha_1}\) expression for \(\abssq{\beta_1}\) and \(P_{\pm}\) [Eq.~\eqref{eq:PpmEN_2}] in Eqs.~\eqref{eq:melEN1}--\eqref{eq:melEN3} we find the normalized density matrix elements up to first order given in the main text [Eqs.~\eqref{eq:melEN_O1-first}--\eqref{eq:melEN_O1-last}].

\section{Optical swaps - single-click}\label{app:optical-swaps}
To better understand the numerical implementation we discuss the optical swaps for the single-click protocol, including the arguments leading to the numerical implementation \cite{CODE} in the following.
The implementation of the double-click protocol follows analogous.
After the initial entanglement generation within the EB and the BB, see Eqs.~\eqref{eq:RHO_BB} and \eqref{eq:RHO_EN}, the next steps are applications of optical swaps to link edge nodes and backbones.
Because we already included all optical losses within the model of the initial generation step, we can apply our detector model to calculate the state after the swap without further losses.
In the absence of dark-counts we find for the first swap (between an EN and a BB link)
\begin{align}
  \label{eq:RHO_SWAP1-non-norm}
  & P_{S1}^{nd} \rho_{S1}^{nd} = 
  \frac{T\abssq{f(t_c)}}{4} \left[ A_0 \proj{0} + A_1' \proj{1} \right] \left[ B_1 + B_1' \right] \proj{\emptyset} 
    + \frac{T\abssq{f(t_c)}}{2} \left[ A_0 \proj{0} + A_1' \proj{1} \right] B_2 2 \proj{1_m}  \\
  & + \frac{T\abssq{f(t_c)}}{2} A_1 \sin^2\theta B_0 \proj{0} \proj{\emptyset} 
    + \frac{T\abssq{f(t_c)}}{4} A_1 B_1 \proj{\varphi_{\pm \sigma \sigma'}} 
    + \frac{T\abssq{f(t_c)}}{4} A_1 B_1' \proj{\varphi_{\mp \sigma \sigma'}} \notag \\
  & + \frac{T\abssq{f(t_c)}}{2} A_1 B_2 \big[ \cos^2\theta \proj{1} \proj{1_m} 
    + (\cos\theta \ket{1} \ket{1_m} \pm \sigma \sigma' \sqrt{2} \sin\theta \ket{0} \ket{2_m}) (\cos\theta \bra{1} \bra{1_m} \pm \sigma \sigma' \sqrt{2} \sin\theta \bra{0} \bra{2_m}) \big] \notag \\
  & + \frac{T\abssq{f(t_c)}}{2} A_2 \proj{1} B_0 \proj{\emptyset} 
    + \frac{T\abssq{f(t_c)}}{4} A_2 \proj{1} \left[ B_1 + B_1' \right] \proj{1_m} 
    + \frac{T\abssq{f(t_c)}}{2} A_2 \proj{1} B_2 2 \proj{2_m} , \notag
\end{align}
where \(\pm\), \(\sigma\), \(\sigma'\) are the ports that clicked in the EN, BB, and swap, respectively.
Here, the states are between the ion \(\ket{0}\) and \(\ket{1}\) and the BB memory \(\ket{k_m}\) (\(k=\emptyset,1,2\)) storing \(k\) photons. 
Additionally, the state \(\ket{\varphi_{\pm}}\) takes the form as in the EN generated state [see Eq.~\eqref{eq:RHO_EN}],
but the memory is the memory of the BB not measured during the swap.
Therefore, this state extends the range of stored matter-photon entanglement.
Note, that within the swaps for simplicity of the numeric implementation we do not disregard mixed higher orders.
Instead, for the dark-counts we again consider heralding vacuum and only account for \(A_0\),\(A_1\),\(B_0\),\(B_1\) terms in agreement with the previous perturbation theory), leading to the additional part
\begin{align}
  \label{eq:RHO-SWAP1-VAC}
  \frac{P^{\text{vac}}_{S1}}{p_d} \rho_{S1}^{\text{vac}} = A_0 B_0 \proj{0} \proj{\emptyset} + \cos^2\theta A_1 B_0 \proj{1} \proj{\emptyset} + \frac{1}{2} A_0 B_1 \proj{0} \proj{1_m} + \frac{\cos^2\theta}{2} A_1 B_1 \proj{1} \proj{1_m} .
\end{align}

We summarize the first swap with the density matrix
in Eq.~\eqref{eq:RHO_SWAP1} \replaced[id=BT]{and}{with}
the (non-normalized) elements \deleted[id=BT]{are}
\begin{align}
  \label{eq:melS1_1}
  P_{S1} C_0 & = \frac{T \abssq{f(t_c)}}{4} \left[ A_0 (B_1 + B_1') + 2  A_1 \sin^2\theta B_0 \right] + p_d A_0 B_0,
  \qquad P_{S1} C_1 = \frac{T\abssq{f(t_c)}}{4} A_1 (B_1 - B_1') , \\
  P_{S1} C_1' & =  \frac{T\abssq{f(t_c)}}{4} \left[ 2 \cos^2\theta A_1 B_1' + A_1' (B_1 + B_1') + 2 A_2 B_0 \right] + p_d \cos^2\theta A_1 B_0 , \\
  P_{S1} C_1'' & = \frac{T\abssq{f(t_c)}}{4} \left[ 2 \sin^2\theta A_1 B_1' + 4 A_0 B_2 \right] + p_d \frac{1}{2} A_0 B_1 , \\
  P_{S1} C_2 & = \frac{T\abssq{f(t_c)}}{4} \left[ 4 A_1' B_2 + 2 \cos^2\theta A_1 B_2 + A_2 (B_1 + B_1') \right] + p_d \frac{\cos^2\theta}{2} A_1 B_1 , \\
  P_{S1} C_2' & = \frac{T\abssq{f(t_c)}}{4} \left[ 2 A_1 B_2 \right] ,
                \qquad P_{S1} C_3 = \frac{T\abssq{f(t_c)}}{4} \left[ 4 A_2 B_2 \right] ,
\end{align}
where all terms apart from \(C_0\) and \(C_1\) have at most a leading order linear in one of the parameters treated as perturbations.
The success probability of the first swap is given by \(\mathbf{P}_{S1} = \int_{\mathbb{S}} dt_c 2 P_{S1}/T\), where \(P_{S1}\) is given by the trace of the unnormalized density matrix \(P_{S1} \rho_{S1}\) and $\mathbb{S}$ is the support of $f$.
For further analysis of our protocol we calculate this density matrix and normalize it numerically, as implemented in Ref.~\cite{CODE}.

Because the state \(\rho_{S1}\) [Eq.~\eqref{eq:RHO_SWAP1}] takes the same form as the edge nodes with the additional elements \(C_1''\), \(C_2'\), and \(C_3\),
we express the final state for both considered approaches (with a central swap or a single backbone link) simultaneously.
To this end we denote both states which are combined by the final swap using $\rho_{S1}$ but substitute  \(C_k \to F_k\) for one of them.
Then $F_k$ takes the values \(A_k\) for an immediate connection of two edge nodes to the same backbone, or \(F_k\) takes the values of \(C_k\) if we employ a swap in the center and thus two backbones (i.e., a multimode repeater).
After the second swap the state takes the form given in the main text [see Eq.~\eqref{eq:final-state}], and the non-normalized elements are
\begin{align}
  \label{eq:melS2_1}
  P_{S2} \alpha & = \frac{T \abssq{f(t_c)}}{2} F_1 C_1 \cos^2\theta \sin^2\theta , \\
  P_{S2} D_{00} & = \frac{T \abssq{f(t_c)}}{2} \left[ C_0 (F_1'' + \sin^2\theta F_1) + (C_1'' + \sin^2\theta C_1) F_0 \right] + p_d C_0 F_0 , \\
  P_{S2} D_{01} & = \frac{T \abssq{f(t_c)}}{2} \left[ (C_1'' + C_1 \sin^2\theta) (F_1' + F_1 \cos^2\theta) + C_0 (F_2 + F_2' \cos^2\theta) \right] + p_d C_0 F_1 \cos^2\theta , \\
  P_{S2} D_{10} & = \frac{T \abssq{f(t_c)}}{2} \left[ (C_1' + C_1 \cos^2\theta) (F_1'' + \sin^2\theta F_1) + (C_2 + C_2' \cos^2\theta) F_0 \right] + p_d C_1 F_0 \cos^2\theta , \\
  P_{S2} D_{11} & = \frac{T \abssq{f(t_c)}}{2} \left[ (C_1' + C_1 \cos^2\theta) (F_2 + \cos^2\theta F_2') + (C_2 + \cos^2\theta C_2') (F_1' + F_1 \cos^2\theta) \right] + p_d F_1 C_1 \cos^4\theta .
\end{align}
The success probability for the second swap is \(\mathbf{P}_{S2} = \int_{\mathbb{S}} dt_c 2 P_{S2}/T\), where \(P_{S2}\) is given by the trace of the unnormalized density matrix \(P_{S2} \rho_{S2}\).

\section{Double-click repeater-less communication rate}\label{app:double-click-no-repeater-rate}

In Sec.~\ref{sec:DC-rate} we discussed the communication rate of the double-click protocol in the presence of a repeater in the center.
Here we provide a discussion of the rate without a repeater.
Without a central repeater we parallelize the generation of both ENs with the sequential generation of two BB links.
Additionally we perform any swap when ready.
Effectively there are two slightly different branches that lead to success: first, we can have a pair of ENs and a single BB link, where the first two swaps are performed in series (geometry) followed by another generation of a BB if the swaps are successful.
We will denote the series swaps success probability as \(\tilde{\mathbf{P}}_{Sk}\) in the following.
Second, we can have an EN and two BB\added[id=BT]{s} and the first two swaps are performed in a parallel geometry (as within one half of the setup for the case with a central repeater).
We will denote the parallel geometry swaps success probability as \(\mathbf{P}_{Sk}\).
In the numerical simulation \cite{CODE}, we ensure that the resulting fidelity of the different branches does not deviate significantly.
For simplicity, we assume that if we first prepare EN-EN and then a BB (or BB-BB and then EN), the first two swaps are attempted simultaneously, such that if we fail either of the swaps, all constituents are reset.

The full average duration is the sum of the probabilities times the average duration to prepare a certain configuration.
The final swaps occur in a parallel geometry if two BBs are prepared first and the first two swaps are successful, i.e., for the preparation sequence BB-BB-EN-EN, BB-EN-BB-EN, and EN-BB-BB-EN,
which occur respectively with probability
\(p_1 =\left( \frac{R_{\text{BB}}}{2R_{\text{EN}} + R_{\text{BB}}} \right)^2\),
\(p_2 = \frac{2 R_{\text{EN}} R_{\text{BB}}}{(2R_{\text{EN}} + R_{\text{BB}})^2} \frac{R_{\text{BB}}}{R_{\text{EN}} + R_{\text{BB}}}\),
\(p_3 = \frac{2 R_{\text{EN}}}{2R_{\text{EN}} + R_{\text{BB}}} \frac{R_{\text{BB}}^2}{(R_{\text{EN}} + R_{\text{BB}})^2}\).
Additionally, they have the associated average preparation duractions
\(T_1 = \left[ \left( \frac{2}{R_{\text{BB}} + 2R_{\text{EN}}} + \frac{1}{2R_{\text{EN}}} \right) \frac{1}{\mathbf{P}_{S1} \mathbf{P}_{S2}} + \frac{1}{R_{\text{EN}}} \right] \frac{1}{\mathbf{P}_{S3} \mathbf{P}_{S4}}\),
\(T_2 = \left[ \left( \frac{2}{R_{\text{BB}} + 2R_{\text{EN}}} \frac{1}{\mathbf{P}_{S1}} + \frac{1}{R_{\text{EN}} + R_{\text{BB}}} \right) \frac{1}{\mathbf{P}_{S2}} + \frac{1}{R_{\text{EN}}} \right] \frac{1}{\mathbf{P}_{S3} \mathbf{P}_{S4}}\), and
\(T_3 = \left\{ \left[  \left( \frac{1}{R_{\text{BB}} + 2R_{\text{EN}}} + \frac{1}{R_{\text{EN}} + R_{\text{BB}}} \right) \frac{1}{\mathbf{P}_{S1}} + \frac{1}{R_{\text{EN}} + R_{\text{BB}}} \right] \frac{1}{\mathbf{P}_{S2}} + \frac{1}{R_{\text{EN}}} \right\} \frac{1}{\mathbf{P}_{S3} \mathbf{P}_{S4}}\).
The final swaps occur in a series geometry, if the first swaps are successful and the constituents are prepared in one of the orders EN-EN-BB-BB, EN-BB-EN-BB, BB-EN-EN-BB, which respectively have the probability
\(p_4 = \frac{2 R_{\text{EN}}}{2R_{\text{EN}} + R_{\text{BB}}} \frac{R_{\text{EN}}}{R_{\text{EN}} + R_{\text{BB}}}\),
\(p_5 = \frac{2 R_{\text{EN}}}{2R_{\text{EN}} + R_{\text{BB}}} \frac{R_{\text{BB}} R_{\text{EN}}}{(R_{\text{EN}} + R_{\text{BB}})^2}\), and
\(p_6 = \frac{2 R_{\text{EN}} R_{\text{BB}}}{(2R_{\text{EN}} + R_{\text{BB}})^2} \frac{R_{\text{EN}}}{R_{\text{EN}} + R_{\text{BB}}}\).
The corresponding average durations are
\(T_4 = \left[ \left( \frac{1}{R_{\text{BB}} + 2R_{\text{EN}}} + \frac{1}{R_{\text{EN}} + R_{\text{BB}}} + \frac{1}{R_{\text{BB}}} \right) \frac{1}{\mathbf{P}_{S1} \tilde{\mathbf{P}}_{S2}} + \frac{1}{R_{\text{BB}}} \right] \frac{1}{\tilde{\mathbf{P}}_{S3} \tilde{\mathbf{P}}_{S4}}\),
\(T_5 = \left\{ \left[  \left( \frac{1}{R_{\text{BB}} + 2R_{\text{EN}}} + \frac{1}{R_{\text{EN}} + R_{\text{BB}}} \right) \frac{1}{\mathbf{P}_{S1}} + \frac{1}{R_{\text{EN}} + R_{\text{BB}}} \right] \frac{1}{\tilde{\mathbf{P}}_{S2}} + \frac{1}{R_{\text{BB}}} \right\} \frac{1}{\tilde{\mathbf{P}}_{S3} \tilde{\mathbf{P}}_{S4}}\), and
\(T_6 = \left[ \left( \frac{2}{R_{\text{BB}} + 2R_{\text{EN}}} \frac{1}{\mathbf{P}_{S1}} + \frac{1}{R_{\text{EN}} + R_{\text{BB}}} \right) \frac{1}{\tilde{\mathbf{P}}_{S2}} + \frac{1}{R_{\text{BB}}} \right] \frac{1}{\tilde{\mathbf{P}}_{S3} \tilde{\mathbf{P}}_{S4}}\).
The full ion-ion generation duration is then given by the sum over the products of probability times average duration of the different sequences \(T_{DC} = \sum_{k=1}^6 p_k T_k\).

\end{widetext}

\section{Direct ion-ion links}\label{app:direct}
In order to have a reference for comparison to our protocol,
we also apply our model to direct ion-ion entanglement \added[id=BT]{generation} using a single and double-click scheme without multiplexing.
In this case, we use the optical setup to directly link two ions from the left \(L\) and right \(R\) which are within a fully symmetric setup.
For the single-click scheme both nodes are described by a state of the form of Eq.~\eqref{eq:ini-atom-SC}.
The non-normalized state (without dark-counts) before tracing over the photon channels is
\begin{align}
  \label{eq:click_state_direct}
  & D_{\pm} \ket{\Psi_L} \ket{\Psi_R}  \\
  & \approx \int_{\mathcal{T}} dt \nu(t) \sqrt{\eta_d} d_{\pm}(t) \left[ \alpha_0 \alpha_1 \ket{\Psi_{\pm}} + \sqrt{1-\eta_d} \alpha_1^2 \ket{1, 1} \right], \notag
\end{align}
with the efficiency \(\eta_d\) which again includes everything from emission up to the click detection.
For direct links the efficiency \(\eta_d\) includes long-range transmission [see Eq.~\eqref{eq:eta_BB}] and potentially frequency conversion.
Analogous to the main text $\alpha_k$ are the $k=0,1$ amplitudes and $\nu$ describes the photon temporal mode, $\ket{\Psi_\pm}$ are the Bell states between the ions and we approximate the dark-counts to herald the no-emission state \(\proj{0,0}\).
Combined we can thus directly calculate the final state in the same form as the fundamental link state within the main text [Eq.~\eqref{eq:final-state}], with the elements
\begin{align}
  \label{eq:direct-ion-1}
  & P \alpha = P D_{0,1} = P D_{1,0} = T \abssq{\nu(t_c)} \frac{\eta_d}{2} \abssq{\alpha_1} (1 - \abssq{\alpha_1}), \\
  & P D_{0,0} = p_d ,
             \quad
   P D_{1,1} = T \abssq{\nu(t_c)} \eta_d (1-\eta_d) \abs{\alpha_1}^4 ,
\end{align}
and click probability \(P = T \abssq{\nu(t_c)} \eta_d \abssq{\alpha_1} \left( 1 -\eta_d \abssq{\alpha_1} \right) + p_d\).
The success probability is again given by summing over the detectors and integrating over the whole pulse, i.e., \(\mathbf{P} = \int_{\mathbb{I}} dt_c 2P/T\) where $\mathbb{I}$ is the support of $\nu$.
In Fig.~\ref{fig:duration}
of the main text we map \(D_{0,0},D_{1,1} \to (D_{0,0} + D_{1,1})/2\) in order to simplify the calculation including an ion repeater.
We note this does not affect the Bell state fidelity of the generated state, and after a ion-ion entanglement swap the fidelity of the mapped state corresponds to the fidelity averaged over the outcomes of the repeater readout.

Analogous, we find for a double-click protocol [see state in Eq.~\eqref{eq:ini-atom-DC}] in the absence of dark counts
\begin{align}
  & D_{\pm} \tilde{D}_{\sigma} \ket{\Psi_L} \ket{\Psi_R} \\
  &\quad = \eta_d \int_{\mathcal{T}'} dt' \int_{\mathcal{T}} dt \nu(t) \nu(t') \frac{d_{\pm,0}(t) \tilde{d}_{\sigma,1}(t')}{2} \ket{\Psi_{\pm \sigma}} . \notag
\end{align}
Thus the ideal click leads to the contribution \(\left( \frac{\eta T }{4} \right)^2 \abssq{\nu(t_c)} \abssq{\nu(t_c')}\) to the non-normalized elements \(P\alpha, P D_{0,1}, P D_{1,0}\)
The dark counts take the same effect for all \(P D_{kl}\) (\(k,l=0,1\)) namely
\(\frac{p_d (1-\eta)}{2} \frac{\eta T [\abssq{\nu(t_c)} + \abssq{\nu(t_c')}]}{4} + \frac{p_d^2 (1-\eta)^2}{4}\).
Here we account for second order dark counts, as the real click probability is affected by both the long pulse duration of the ion and the long distance fiber losses simultaneously.
Note that in the numerical simulation we take \(\abssq{\nu(t_c)} = {1}/{T_a}\) as constant over the support domain \([0,T_a]\) (and \(0\) else).

\begin{figure*}[ht]
\centering
{
\includegraphics[width=\linewidth]{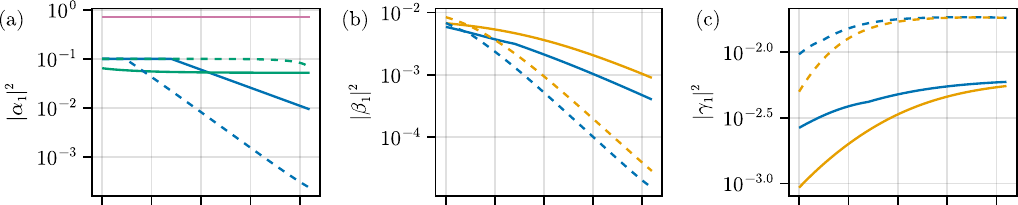} \\
\includegraphics[width=\linewidth]{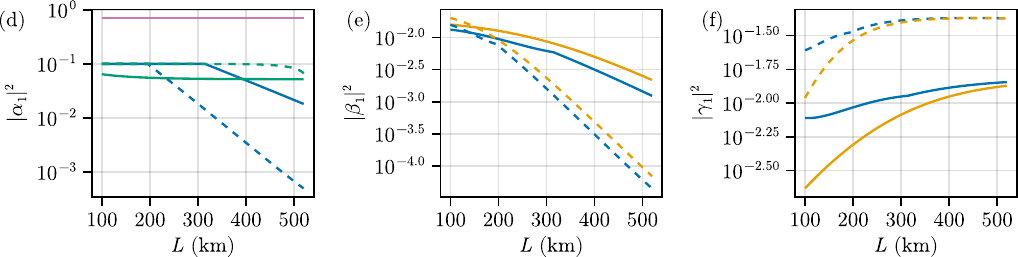}
}
\caption{\label{fig:probs_lower_effs} \label{fig:probs_higher_effs}
  Optimized emission probabilities corresponding to Fig.~\ref{fig:duration}, where we provide the used parameters.
  Here, (a-c) correspond to the memory efficiency $\eta_m=0.5$ and (d-f) to $\eta_m=0.8$.
  The line style encodes the protocols as explained in Fig.~\ref{fig:duration}.
  In panel (a,d) we show the emission probability of the ions \(\abssq{\alpha_1}\), in (b,e) of the SPDC connecting memory and ions in the edge nodes \(\abssq{\beta_1}\) and in (c,f) of the SPDC within the BB \(\abssq{\gamma_1}\). 
}
\end{figure*}

\section{Deterministic ion-ion entanglement swap}\label{app:ion-swaps}
As we employ a central multimode repeater in parts of the analysis, we also calculate the effect of direct ion-ion swapping for comparison with the direct ion-ion links.
Note, that we furthermore propose that our fundamental links can be extended using these ion-ion swaps.
We begin from a set of two \replaced[id=BT]{ion-ion}{atom-atom} entangled links (in series geometry) that each take the form given in Eq.~\eqref{eq:final-state}.
The two ions in the center have access to two-qubit gates, e.g., by sharing the same trap,
such that we perform a CNOT gate followed by a Hadamard gate on the control bit, this maps the shared state according to
\begin{align}
  \label{eq:swap}
  \ket{k,l} \otimes \ket{\tilde{k}, \tilde{l}} \to \frac{1}{\sqrt{2}} \left( \ket{k,0} + (-1)^l \ket{k,1} \right) \otimes \ket{\abs{\tilde{k} - l}, \tilde{l}} .
\end{align}
Then the repeater (central) qubits are measured in the Z-basis.
If the central qubit belonging to the left link is measured in \(\ket{1}\) we apply a Pauli \(Z\)-gate to the left most qubit and if the central qubit of the right link was read out in \(\ket{0}\) we apply a Pauli \(X\)-gate on the right qubit.
If the initial states satisfy \(\alpha>0\), \(D_{0,1} = D_{1,0}\), and \(D_{1,1} = D_{0,0}\), we are then left with a state of the same form as the initial links for the outer pair, but with \(\alpha \to {\alpha^2}/{2}\), \(D_{0,1} \to (D_{0,1}^2 + D_{0,0}^2)/2\), and \(D_{0,0} \to D_{0,1} D_{0,0}\).
As all possible measurement results of the qubits in the center lead to a successful swap the process is deterministic,
and for \(D_{0,0} \ll D_{1,0}\) the error is approximately doubled during the swap since an error in either of the entangled states results in an error after the swap.

\section{Additional details on the optimization}\label{app:opt}

In Fig.~\ref{fig:duration_lower_effs} of the main text we presented the average duration to prepare a state for a fixed fidelity after optimizing over the emission probabilities.
In this appendix we provide further details on the optimization.
The emission probabilities corresponding to the results in Fig.~\ref{fig:duration_lower_effs}
of the main text are displayed in Fig.~\ref{fig:probs_lower_effs}.
For the protocol proposed in the main text we numerically optimize the emission probabilities
\(\abssq{\alpha_1}, \abssq{\beta_1},\) and \(\abssq{\gamma_1}\)
using our numerical implementation \cite{CODE} and \enquote{Optim.jl} \cite{mogensen-2018-julia_optim}.
\added[id=BT]{
We use a constrained optimization using the \enquote{IPNewton()} method of \enquote{Optim.jl}, where derivatives are evaluated using automatic forward differentiation.
The constraints are that the fidelity is larger than the target fidelity, as well as the (numerically optimized) emission probabilities are in between $10^{-12}$ and $10^{-1}$.
The upper bound of the emission probabilities ensures validity of the perturbatively derived states of the initial entanglement generation and explains why the emission probabilities saturate in Fig.~\ref{fig:probs_lower_effs}.
The initial guess for the optimization is determined for the first data point (shortest distance) by using a fixed relation between the emission probabilities, and then gradually reducing them until they satisfy the fidelity bound (for more details see \cite{LETT,CODE}, the remaining data points reduce the emission probabilities by a factor to get the guess for the next point.
This factor needs to be chosen manually dependent on the remaining parameters.
}

\added[id=BT]{Some of the protocols need to account for particular details, which we detail below.}
Note that for the double-click protocol \(\alpha_1 = {1}/{\sqrt{2}}\) is not an optimization parameter and for the direct ion-ion generation there are no SPDC emission amplitudes \(\beta_1\) and \(\gamma_1\).
We additionally note that for the single-click direct ion-ion approach we use \(10000\) exponentially spaced samples for the ion emission probability \(\abssq{\alpha_1}\) between \(10^{-1}\) and \(10^{-6}\) decreasing until we find the first value that satisfies the fidelity constraint.
The ion-ion double-click protocol, does not need any optimization, as both ions each emit a photon shared evenly between the two rails.

\added[id=BT]{
Note, that given the more deterministic approaches for the direct ion-ion entanglement generation, we expect the conclusions of the numerical optimization for the hybrid approach to be valid even if results in a local optimum.
}
\added[id=AS]{
Any further optimization can only improve the performance of the hybrid approach introduced here and it will thus have an even larger advantage.
}

\bibliography{refs.bib}

\end{document}